\documentclass[a4paper,11pt]{article}
\usepackage{latexsym,amsmath,amsfonts,amssymb}
\usepackage[latin1]{inputenc}
\usepackage[american]{babel}
\usepackage[dvips]{graphicx}
\usepackage{bbm}

\newcommand{\Vol}{\mathrm{Vol}}

\numberwithin{equation}{section}

\newcommand{\Tr}{\mbox{Tr}}    

\newcommand{\be}{\begin{equation}} \newcommand{\ee}{\end{equation}}
\newcommand{\bea}{\begin{equation} \begin{aligned}} \newcommand{\eea}{\end{aligned} \end{equation}}

\newcommand{\calI}{\mathcal{I}}

\newcommand{\calN}{\mathcal{N}}
\newcommand{\calO}{\mathcal{O}}

\newcommand{\bbC}{\mathbb{C}}

\newcommand{\bbP}{\mathbb{P}}
\newcommand{\bbR}{\mathbb{R}}
\newcommand{\bbZ}{\mathbb{Z}}

\begin{document}

\makeatletter \@addtoreset{equation}{section} \makeatother
\renewcommand{\theequation}{\thesection.\arabic{equation}}
\pagestyle{empty}

\rightline{TAUP-2917/10}
\vspace{2.5cm}
\begin{center}
{\LARGE{\bf Type IIB construction of flavoured ABJ(M)\\ and fractional M2 branes \\[20mm]}} {\Large{Stefano Cremonesi}}
\bigskip

{\small{{}
Raymond and Beverly Sackler School of Physics and Astronomy \\
 Tel-Aviv University, Ramat-Aviv 69978, Israel\\
\medskip \tt{stefano@post.tau.ac.il}
}}
\bigskip

\bigskip

\bigskip

\bigskip

\bigskip

{\bf Abstract}
\vskip 20pt
\end{center}
We study type IIB brane configurations engineering 3d flavoured ABJ(M) theories with Yang-Mills kinetic terms, which flow to IR fixed points describing M2 branes at a class of toric Calabi-Yau fourfold singularities.
The type IIB construction provides a bridge between M-theory geometry and field theory, and allows to identify the superconformal field theories with fixed quiver diagram, Chern-Simons levels and superpotential, differing by the ranks of the gauge groups, which we associate to dual $AdS_4\times Y_7$ backgrounds of M-theory without or with torsion $G$-fluxes sourced by fractional M2 branes in $Y_7$, when $Y_7$ is smooth. The analysis includes the $Q^{1,1,1}$ and $Y^{1,2}(\mathbb{CP}^2)$ geometries.
We also comment on duality cascades and on the interplay between torsion $G$-fluxes in M-theory and partial resolutions.

%
%
\begin{center}
\begin{minipage}[h]{16.0cm}

\end{minipage}
\end{center}
\newpage
\setcounter{page}{1} \pagestyle{plain}
\renewcommand{\thefootnote}{\arabic{footnote}} \setcounter{footnote}{0}

\tableofcontents

\vspace*{1cm}


\section{Introduction}\label{sec:Intro}

Following the work of Aharony, Bergman, Jafferis and Maldacena (ABJM) \cite{Aharony:2008ug}, who introduced a quiver gauge theory with $U(N)\times U(N)$ gauge group and CS terms at levels $\pm k$ to describe the low energy dynamics of a stack of M2 branes at $\bbC^4/\bbZ_k$ and made several checks of the duality between such a three-dimensional (3d) field theory and M-theory on $AdS_4\times S^7/\bbZ_k$, a wealth of publications  appeared generalising the correspondence to systems of M2 branes at singularities, preserving less supersymmetry.

Of particular interest to us is the case of M2 branes at conical toric Calabi-Yau fourfolds ($CY_4$), where the field theory has $\calN=2$ supersymmetry in three dimensions and at least $U(1)_R\times U(1)^3$ global symmetry, and whose study was initiated in \cite{Martelli:2008si,Hanany:2008cd} and \cite{Ueda:2008hx,Imamura:2008qs}, where the connection to the M-theory brane crystal proposal of  \cite{crystals}  was further investigated. It was proposed that given a four-dimensional (4d) quiver gauge theory for D3 branes at a toric $CY_3$ cone, the same toric quiver gauge theory, reinterpreted as a three-dimensional field theory with CS terms, could be useful as a description of the low energy dynamics on M2 branes at certain toric $CY_4$ singularities. The toric fourfold is the total space of a line bundle over the threefold, where the fibration is related to the CS levels of the 3d gauge theory (see also \cite{Jafferis:2008qz,Aganagic:2009zk}), which are subject to the constraint that their sum vanishes. Such three dimensional field theories are known as 3d quiver gauge theories with a 4d parent. 

3d quiver gauge theories without 4d parent, inspired by M-theory brane crystals, were also proposed in \cite{Hanany:2008cd} and further investigated in \cite{Hanany:2008fj,Franco:2008um} and references thereof. These are quiver gauge theories which in 4d do not lead to fixed points and are believed not to describe stable D3 branes at toric $CY_3$ cones. 
It was noted in \cite{Benini:2009qs,Jafferis:2009th} that in such cases the fibration has fixed points, therefore reducing to type IIA along the circle of the fibre leads to D6 branes at the noncompact fixed point loci. The observation led to the proposal that 3d quiver gauge theories without 4d parents should be replaced by 3d quiver gauge theories with 4d parent, further endowed with fundamental and antifundamental flavours arising from D2-D6 and D6-D2 strings. The flavour proposal correctly reproduces the abelian moduli space like the older one, but has a number of advantages: it has a string theory derivation, it makes manifest the nonabelian flavour symmetries associated to non-isolated singularities, and thus it accounts for their partial resolutions.

This article focuses on a simple yet interesting subclass of such flavoured toric quiver gauge theories, namely flavoured ABJM models, which allows a dual construction using systems of D3 branes suspended between fivebranes in type IIB string theory. The benefit ot this approach, that can only be applied to a subset of the 3d flavoured quiver CS gauge theories studied in \cite{Benini:2009qs}, is that basic laws of brane physics, such as the $s$-rule and the D3 brane creation effect \cite{Hanany:1996ie}, can be used to study 3d Seiberg-like dualities and fractional M2 branes in M-theory, as in \cite{Aharony:2008gk}. In addition, the connection between the flavoured 3d toric quiver gauge theories and their toric $CY_4$ geometric moduli spaces, which are metric cones over toric Sasaki-Einstein 7-folds $Y_7$, is apparent in the dual type IIB brane configurations.
  
This analysis allows us to identify, by means of the $s$-rule of brane dynamics in type IIB string theory, the inequivalent 3d superconformal field theories of flavoured ABJ(M) type with given superpotential, CS levels, number of fundamental flavours and smaller common gauge rank. These are expected to be dual to $AdS_4\times Y_7$ backgrounds of M-theory with different torsion $G$-fluxes in $Y_7$, because only torsion $G$-fluxes do not spoil the $AdS_4\times Y_7$ structure of the background when turned on \cite{Aharony:2008ug,Benishti:2009ky}.
We identify the 3d SCFTs dual to the $AdS_4\times Y_7$ solutions with and without torsion fluxes for two notable $Y_7$ geometries that have been studied in the literature, namely $Q^{1,1,1}$ and $Y^{1,2}(\mathbb{CP}^2)$. 
A similar strategy was pursued in \cite{Imamura:2008ji}, where brane configurations preserving twice the amount of supersymmetry were considered.

The type IIB result applies to each representative of the 5-parameter family of flavoured ABJ(M) theories. Following ABJ \cite{Aharony:2008gk}, one may think that the number $n$ of inequivalent 3d superconformal field theories of flavoured ABJ(M) type with given superpotential, CS levels and number of flavours computed in IIB corresponds to the order of the torsion part of $H^4(Y_7,\bbZ)$ for any $Y_7$ appearing in the family of geometries under consideration. 
There are however two partially related caveats to this. As we will explain, subtleties  appear if $Y_7$ is singular, therefore we will mainly focus on smooth $Y_7$, two one-parameter families of which appear in our construction. Secondly, one may wonder that our type IIB/field theory analysis might miss some inequivalent SCFT's with the same geometric moduli space: in such a case, the torsion fourth cohomology group would only include the $\bbZ_n$ expected on the ground of the type IIB analysis. For reasons explained in section \ref{subsec:fracM2}, we expect the latter subtlety to appear only in the presence of singularities in $Y_7$, and we conjecture that for smooth Sasaki-Einstein geometries $H^4_{tor}(Y_7,\bbZ)=\bbZ_n$.
In addition to the matching found for the $Q^{1,1,1}$ and $Y^{1,2}(\mathbb{CP}^2)$ geometries, we extend the computation of $H^4_{tor}(Y^{1,2}(\mathbb{CP}^2),\bbZ)$ in \cite{Benishti:2009ky} to a one-parameter family of smooth $Y_7$ which includes $Y^{1,2}(\mathbb{CP}^2)$, and check that the type IIB brane computation of $n$ and the algebraic topology computation of $H^4(Y_7,\bbZ)$ match in such a class, in agreement with the physical expectation.

The paper is organised as follows. In section \ref{sec:vectorlike} we introduce the type IIB construction for the three-dimensional version of the Klebanov-Witten theory \cite{Klebanov:1998hh}, first without and then with vectorlike doubled flavours, using D3 branes suspended on a circle between NS5 branes, possibly intersecting D5 branes. In section \ref{sec:flavoured_ABJ} we review how web deformations of the fivebrane systems, which are interpreted as real mass terms for fundamental flavours in the 3d gauge theory on the noncompact dimensions of the D3 branes, may be used to find brane configurations for flavoured ABJ(M) theories, and relate the fivebrane webs in type IIB to the toric diagram of the Calabi-Yau fourfold probed by the M2 branes in M-theory.
In section \ref{subsec:HWeffect_flavoured_ABJM} we study the D3 brane creation effect that occurs if the two fivebranes cross each other when sliding in the mutually transverse circle direction. This effect and the $s$-rule determine the numerology of duality cascades and the superconformal fixed points of flavoured ABJ theories with different gauge ranks, which are conjectured to be related to fractional M2 branes that are M5 branes wrapped on torsion 3-cycles in M-theory. We provide a check of this conjecture for a one-parameter family of smooth Sasaki-Einstein 7-folds appearing in the present construction, in addition to the examples of $Q^{1,1,1}$ and $Y^{1,2}(\mathbb{CP}^2)$, for which the cohomologies were computed in the literature.
In section \ref{sec:Higgsing} we study the interplay between fractional M2 branes and partial resolutions of the Calabi-Yau singularity from the viewpoint of the type IIB dual. We conclude in section \ref{sec:conclusions} with a summary and an outlook. Finally, appendix \ref{app:volumes} collects the volumes of two families of smooth Sasaki-Einstein 7-folds that are introduced in the main body of the paper, as well as those of their supersymmetric 5-cycles.


\section{Type IIB brane constructions for flavoured 3d KW theories}\label{sec:vectorlike}

In this section we first review how 3d versions of the Klebanov-Witten (KW) quiver gauge theories are engineered using D3 branes suspended between NS5 branes in type IIB string theory, and then we explain how the type IIB construction accounts for the introduction of vectorlike doubled flavours.

\subsection{3d Klebanov-Witten theory}\label{subsec:3dKW}

Our starting point is a configuration of D3 branes suspended between NS5 branes in type IIB string theory, that engineers a $U(N)\times U(N)$ 3d Yang-Mills (YM) quiver gauge theory with bifundamental matter fields and quartic superpotential, as in its four-dimensional counterpart analysed by Klebanov and Witten  \cite{Klebanov:1998hh} in the context of the $AdS_5/CFT_4$ correspondence. The Klebanov-Witten field theory for D3 branes at the conifold was later related in \cite{Uranga:1998vf,Dasgupta:1998su} to a T-dual type IIA brane configuration of D4 branes on a circle intersecting two NS5 branes with a relative angle. The type IIB brane construction that we are about to review is related to the one of \cite{Uranga:1998vf,Dasgupta:1998su} by a T-duality along a field theory direction. We will generalise to different ranks later on.

The setting is type IIB string theory on $\bbR^{1,8}\times S^1$, $x^6$ being a coordinate along the circle of radius $R_6$: $x^6\sim x^6+2\pi R_6$.
Let us initially consider the following brane configuration: 
\begin{itemize}
\item an NS5 along $012345$ at $x^6=x^7=x^8=x^9=0$;
\item an NS5$_\theta$ along $0123[48]_\theta [59]_\theta$ at $x^6=2\pi R_6\cdot b$, $x^7=0$, $(x^4+i x^5) \sin\theta = (x^8+i x^9) \cos\theta$;
\item $N$ D3 branes along $0126$, wrapping the circle and at $x^4=x^5=x^7=x^8=x^9=0$.
\end{itemize}
When $\theta=0$, the two NS5 branes are parallel and the low energy 3d field theory on the noncompact dimensions of the D3 branes is the $\calN=4$ $U(N)\times U(N)$ gauge theory with Kronheimer necklace quiver with two nodes, in the eight supercharge notation where arrows represent hypermultiplets. Each gauge group comes with an $\calN=4$ vector multiplet with YM kinetic term. They decompose into $\calN=2$ vector multiplets and $\calN=2$ chiral multiplets, $\Phi$ transforming in $(adj,1)$ and $\tilde\Phi$ in $(1,adj)$. Moreover there are two bifundamental hypermultiplets corresponding to 3-3 strings passing through each of the solitonic fivebranes: they decompose into $(A_1, B_1^\dagger)$ in $(\square,\overline{\square})$, and $(B_2, A_2^\dagger)$ in $(\overline{\square},\square)$, $A_i$ and $B_j$ being $\calN=2$ chiral multiplets. This brane configuration is T-dual (along $x^6$) to $N$ D2 branes probing the $A_1$ singularity, together with a transverse $\bbR^3$. 

When $\theta\neq 0$, the adjoint chiral superfields in the $\calN=4$ vector multiplets acquire equal and opposite mass $m$, and the superpotential is%
\footnote{In our convention gauge couplings only appear in front of the YM kinetic terms in the action, so that both the gauge bosons and the chiral adjoints $\Phi$ and $\tilde\Phi$ have engineering dimension 1. Therefore $m$ is dimensionless here, or, in other words, it is measured in units of the squared YM coupling.}
\begin{equation}\label{W_before_integrating_out}
W = \sqrt{2}\, \Tr\left[\Phi(A_1 B_1-A_2 B_2)\right] -  \sqrt{2}\, \Tr\left[\tilde\Phi(B_1 A_1-B_2 A_2)\right] + \frac{m}{2}\Tr(\Phi^2)-\frac{m}{2}\Tr(\tilde\Phi^2)\;.
\end{equation}
Integrating out the massive adjoints, the low energy superpotential is quartic
\begin{equation}\label{W_after_integrating_out}
W = h\,\Tr(A_1B_1A_2B_2-A_1B_2A_2B_1)= h\, \epsilon^{ij}\epsilon^{kl} \,\Tr(A_iB_kA_jB_l)
\end{equation}
with $h=\frac{2}{m}$, and preserves an $SU(2)\times SU(2)$ non-R symmetry that is not manifest in the brane picture, in addition to the $U(1)_R$ symmetry. Henceforth, we will refer to the three-dimensional quiver gauge theory with the matter content previously introduced and superpotential \eqref{W_after_integrating_out} as the three-dimensional Klebanov-Witten theory (3d KW).

For small angles, $m\propto \tan\theta$, but for large angles such a trigonometric relation is believed to receive corrections, so that a quartic superpotential superpotential with $h\neq 0$ survives even for fivebranes at right angles. This is required in order to reproduce the cubic superpotential of $\calN=8$ SYM upon diagonal Higgsing $U(N)\times U(N)\to U(N)$ in the presence of a Fayet-Iliopoulos (FI) term \cite{Uranga:1998vf}.

\begin{figure}[t]
\begin{center}
\includegraphics[width=12cm]{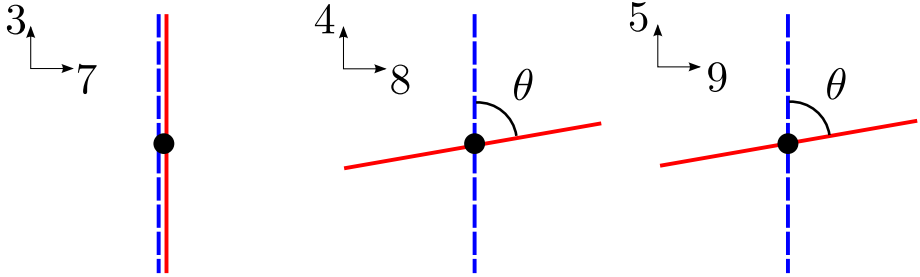}
\caption{\small Type IIB brane configuration for the 3d KW theory. The two NS5 branes are depicted by (dashed) blue and red lines, the suspended D3 branes by the black dot. The $x^6$ circle is suppressed.}\label{fig:3d_KW_1}
\end{center}
\end{figure}
The type IIB brane configuration for 3d KW is depicted in figure \ref{fig:3d_KW_1}. The two NS5 branes are separated along the circle direction $x^6$, which is orthogonal to the page. 
The inverse squared Yang-Mills couplings of the two gauge groups are proportional to the lengths of the $x^6$ arcs between the fivebranes, and therefore to $b$ and $1-b$, with $b\in[0,1]$.
Setting our conventions, we associate the bifundamental fields $A_1$ and $B_1$ to 3-3 strings intersecting the blue (dashed) NS5 brane (with the two orientations respectively), and $B_2$, $A_2$ to 3-3 strings intersecting the red NS5$_\theta$ brane.  
This type IIB brane configuration is T-dual (along $x^6$) to $N$ D2 branes probing the conifold singularity together with a transverse $\bbR$ in type IIA.


\subsection{3d KW theory with vectorlike doubled flavours}\label{subsec:flavoured_KW}

We now proceed to add vectorlike \emph{doubled}%
\footnote{We prefer to refer to the $SU(N_f)\times SU(N_f)$ flavour symmetry of 3d $\calN=2$ SQCD with vectorlike flavours as \emph{doubled} rather than \emph{chiral}, since there is no chirality in three spacetime dimensions.}
flavours as suggested in \cite{Brodie:1997sz}. See also \cite{Aharony:1997ju,Ouyang:2003df} and in particular \cite{Brunner:1998jr} for a careful account. 
To this aim, we decorate the brane configuration presented in the previous subsection (with  generic $\theta\neq 0$) with:
\begin{itemize}
\item $F_1$ D5 along $012457$ at $x^3=x^6=x^8=x^9=0$;
\item $F_2$ D5$_\theta$ along $012[48]_\theta [59]_\theta 7$ at $x^6=2\pi R_6\cdot b$, $x^3=0$, $(x^4+i x^5) \sin\theta = (x^8+i x^9) \cos\theta$.
\end{itemize}
The $F_1$ D5 branes along $012457$ are split into two halves by the NS5 brane at $x^6=0$, therefore they contribute a $U(F_1)^2$ flavour symmetry to the low energy three-dimensional field theory on the noncompact dimensions of the suspended D3 branes.%
\footnote{To be precise, these are the global flavour symmetries if the gauge dynamics is turned off. At non-vanishing gauge couplings, an overall diagonal $U(1)$ subgroup is not a global symmetry since it acts like the diagonal $U(1)$ gauge symmetry.} 
Similarly, the $F_2$ D5$_\theta$ branes are split into two halves by the NS5$\theta$ brane at $x^6=2\pi R_6\cdot b$, thereby providing a $U(F_2)^2$ flavour symmetry.

\begin{figure}[t]
\begin{center}
\includegraphics[width=11cm]{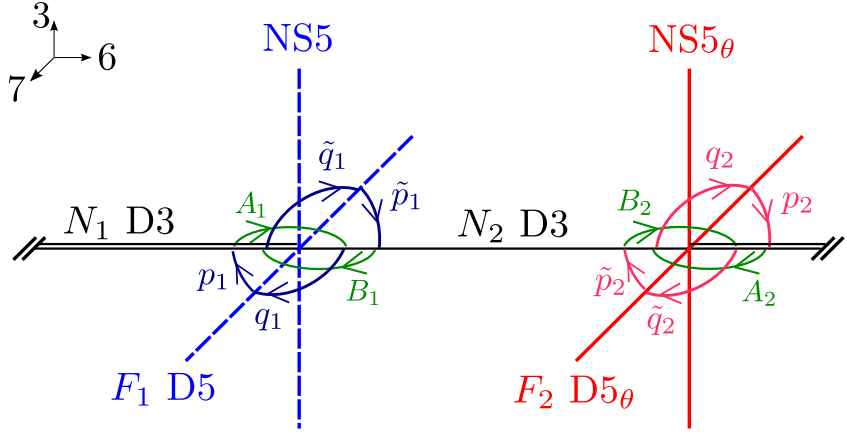}
\end{center}
\caption{\small Type IIB brane configuration, open strings and matter fields (4589 directions suppressed).}\label{fig:chiral_flavours_x6}
\end{figure}

It may be useful to visualise the bifundamental and fundamental matter content in terms of open strings as in fig. \ref{fig:chiral_flavours_x6}.
The resulting matter content of the flavoured gauge theory, allowing for generic ranks of the gauge group $U(N_1)\times U(N_2)$ and with vectorlike flavour group $\left[U(F_1)^2\times U(F_2)^2\right]/U(1)_{diag}$, is summarised in the following table:
\begin{center}
\begin{tabular}{c|c c | c c c c}
             & $U(N_1)$  & $U(N_2)$  & $U(F_1)$ & $U(F_1)$ & $U(F_2)$ & $U(F_2)$  \\ \hline
$A_1$, $A_2$ & $\square$ & $\overline{\square}$  & $1$   &  $1$ &  $1$ &  $1$   \\
$B_1$, $B_2$ & $\overline{\square}$ & $\square$  & $1$   &  $1$ &  $1$ &  $1$   \\
\hline
$p_1$        & $\overline{\square}$ & $1$ & $\square$ &  $1$ &  $1$ &  $1$   \\
$q_1$        & $1$ & $\square$ & $\overline{\square}$ &  $1$ &  $1$ &  $1$   \\ \hline
$\tilde p_1$ & $1$ & $\overline{\square}$ &  $1$ &  $\square$ & $1$ &  $1$   \\
$\tilde q_1$ & $\square$ & $1$ &  $1$ & $\overline{\square}$ &   $1$ & $1$   \\ \hline
$p_2$        & $\overline{\square}$ & $1$  &  $1$ &  $1$ & $\square$ &  $1$  \\
$q_2$        & $1$ & $\square$ &  $1$ &  $1$ & $\overline{\square}$ &  $1$   \\ \hline
$\tilde p_2$ & $1$ & $\overline{\square}$ & $1$ &  $1$ &  $1$ &  $\square$   \\
$\tilde q_2$ & $\square$ & $1$ &   $1$ & $1$ & $1$ & $\overline{\square}$   
\end{tabular}\label{fig:matter_content_vectorlike}
\end{center}
\vspace{5pt}
Assuming $\theta \neq 0$ and reabsorbing couplings, the low energy superpotential can be written as
\begin{equation}\label{W_flavoured_KW_vectorlike}
\begin{split}
W &= \Tr(A_1B_1A_2B_2-A_1B_2A_2B_1) \,+\\
&+ \sum_{\alpha=1}^{F_1} (p_1)^\alpha A_1 (q_1)_\alpha - \sum_{\beta=1}^{F_1} (\tilde p_1)^\beta B_1 (\tilde q_1)_\beta -  \sum_{\gamma=1}^{F_2} (p_2)^\gamma A_2 (q_2)_\gamma + \sum_{\delta=1}^{F_2} (\tilde p_2)^\delta B_2 (\tilde q_2)_\delta \;,
\end{split}
\end{equation}
where contraction of gauge indices is implied in the second line. We will refer to this quiver gauge theory, with YM kinetic terms, as the 3d KW theory with $F_1+F_2$ doubled vectorlike flavours.

\begin{figure}[t]
\begin{center}
\includegraphics[width=12cm]{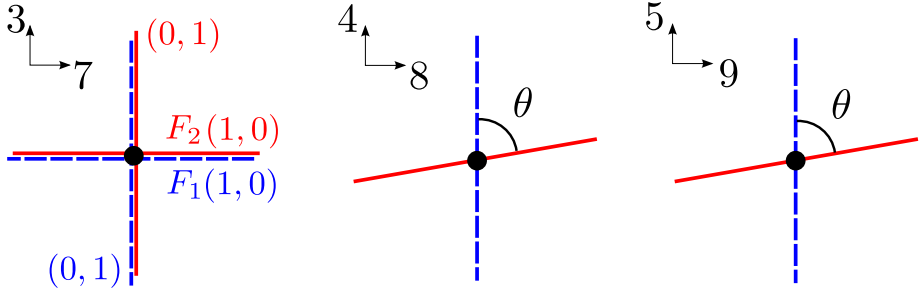}
\caption{\small Type IIB brane configuration for 3d KW with $F_1+F_2$ doubled vectorlike flavours. }\label{fig:3d_KW_1_Ouyang_flav}
\end{center}
\end{figure}

Finally, we draw in fig. \ref{fig:3d_KW_1_Ouyang_flav} the type IIB brane configuration for the 3d KW theory with $F_1+F_2$ vectorlike flavours, in a way that allows comparison to fig. \ref{fig:3d_KW_1} for the unflavoured 3d KW theory. 
From now on the 45 and 89 planes will be suppressed in the figures. The brane embeddings in such planes is understood to remain as in figures \ref{fig:3d_KW_1} and \ref{fig:3d_KW_1_Ouyang_flav} (with generic $\theta\neq 0$) unless otherwise stated.


\section{Toric flavoured ABJ(M) theories}\label{sec:flavoured_ABJ}

In this section we explain how web deformations of the fivebrane systems, interpreted as real mass deformations in the 3d gauge theory, allow us to obtain type IIB brane configurations for quiver gauge theories with the bifundamental matter content of KW, fundamental and antifundamental matter charged under the flavour group $[U(h_a)\times U(h_b)\times U(h_c)\times U(h_d)]/U(1)$, and possibly $\calN=2$ Chern-Simons (CS) interactions compatible with gauge invariance.%
\footnote{See \cite{Bergman:1999na,Aharony:2008ug} for earlier discussions where the resulting field theories lack fundamental flavours.}
\begin{figure}
\begin{center}
\includegraphics[width=6cm]{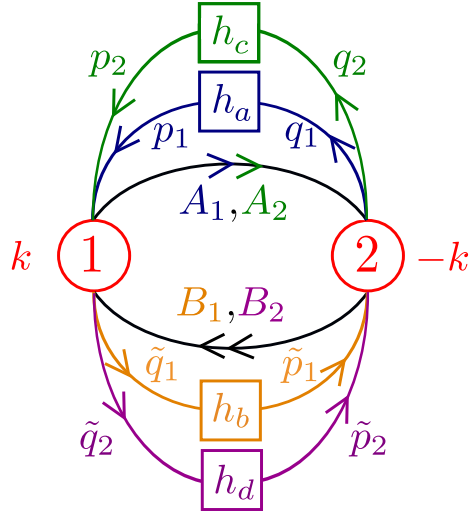}
\caption{\small Quiver diagram of flavoured ABJM theories: circles are gauge groups, squares are flavour groups, arrows are chiral superfields in the fundamental representation of the tail and antifundamental representation of the head. CS levels are denoted in red.}\label{fig:KW4flav}
\end{center}
\end{figure}
The matter content is summarised in the quiver diagram of fig. \ref{fig:KW4flav} and transforms in the following representations:
\begin{center}
\begin{tabular}{c|c c | c c c c}
             & $U(N_1)$  & $U(N_2)$  & $U(h_a)$ & $U(h_b)$ & $U(h_c)$ & $U(h_d)$  \\ \hline
$A_1$, $A_2$ & $\square$ & $\overline{\square}$  & $1$   &  $1$ &  $1$ &  $1$   \\
$B_1$, $B_2$ & $\overline{\square}$ & $\square$  & $1$   &  $1$ &  $1$ &  $1$   \\
\hline
$p_1$        & $\overline{\square}$ & $1$ & $\square$ &  $1$ &  $1$ &  $1$   \\
$q_1$        & $1$ & $\square$ & $\overline{\square}$ &  $1$ &  $1$ &  $1$   \\ \hline
$\tilde p_1$ & $1$ & $\overline{\square}$ &  $1$ &  $\square$ & $1$ &  $1$   \\
$\tilde q_1$ & $\square$ & $1$ &  $1$ & $\overline{\square}$ &   $1$ & $1$   \\ \hline
$p_2$        & $\overline{\square}$ & $1$  &  $1$ &  $1$ & $\square$ &  $1$  \\
$q_2$        & $1$ & $\square$ &  $1$ &  $1$ & $\overline{\square}$ &  $1$   \\ \hline
$\tilde p_2$ & $1$ & $\overline{\square}$ & $1$ &  $1$ &  $1$ &  $\square$   \\
$\tilde q_2$ & $\square$ & $1$ &   $1$ & $1$ & $1$ & $\overline{\square}$   
\end{tabular}\label{fig:matter_content}
\end{center}
\vspace{5pt}
The flavour groups $U(h_a)\times U(h_b)$ and $U(h_c)\times U(h_d)$ are subgroups of $U(F_1)^2$ and $U(F_2)^2$ respectively, which remain as global symmetries after the infinite real mass deformation.
Absorbing couplings, the superpotential is \cite{Benini:2009qs}
\begin{equation}\label{W_flavoured_KW}
\begin{split}
W &= \Tr(A_1B_1A_2B_2-A_1B_2A_2B_1) \,+\\
&+ \sum_{\alpha=1}^{h_a} (p_1)^\alpha A_1 (q_1)_\alpha - \sum_{\beta=1}^{h_b} (\tilde p_1)^\beta B_1 (\tilde q_1)_\beta -  \sum_{\gamma=1}^{h_c} (p_2)^\gamma A_2 (q_2)_\gamma + \sum_{\delta=1}^{h_d} (\tilde p_2)^\delta B_2 (\tilde q_2)_\delta \;.
\end{split}
\end{equation}
The CS levels are $\vec{k}\equiv(k_1,k_2)=(k,-k)$ are half-integers or integers depending on whether or not the the matter content induces a parity anomaly. Indeed, invariance of the partition function under large gauge transformations requires the quantisation law
\be\label{level_quantisation}
\pm k + \frac{1}{2} (h_a-h_b+h_c-h_d) \,\in \bbZ \;.
\ee
We will refer to this class of theories as flavoured ABJM models, depending on five parameters: the CS level and the number of flavours coupled trilinearly to each of the four bifundamentals.

As anticipated, the field theories introduced in this section may be obtained from those of section \ref{subsec:flavoured_KW} by giving real masses to some chiral multiplets charged under the flavour group, and integrating them out.
Real mass terms in three dimensions have a supersymmetric description as D-terms of the form
\begin{equation}\label{real_mass_SUSY}
\int d^4\theta\, Z^\dagger \,e^{m \, \theta\bar{\theta}}\, Z\;.
\end{equation}
Recall that, since the minimal coupling of the chiral superfield $Z$ to a vector superfield $V$ is
\begin{equation}\label{minimal_coupling}
\int d^4\theta\, \Tr \left(Z^\dagger \,e^V\, Z\right)
\end{equation}
and $V\supset \sigma\, \theta\bar{\theta}$, where $\sigma$ is the noncompact real scalar in the vector multiplet, the real mass matrix receives contributions from the vacuum expectation values (VEV) of the real scalar $\sigma$, both if $\sigma$ is the dynamical field of a gauge vector multiplet or if it is the spurion of a background vector multiplet associated to a global symmetry \cite{Aharony:1997bx}.

Specifically, we are interested in giving to the eigenvalues of the hermitian adjoint spurions of the four flavour groups $\sigma_a$, $\sigma_b$, $\sigma_c$ and $\sigma_d$ respectively $F_1-h_a$, $F_1-h_b$, $F_2-h_c$ and $F_2-h_d$ nonvanishing VEVs.
As a result, some of the flavour fields acquire a real mass. At energies smaller than the absolute value of the real mass $M_\psi$ of a chiral multiplet $\psi$, the massive multiplet is integrated out. Integrating out massive fermions generates an effective CS term at one loop.
For definiteness, let us consider the symmetry breaking $U(F_1)\to U(h_a)\times U(1)^{F_1-h_a} $ by means of a spurion $\langle \sigma_a\rangle=\mathrm{diag}(0,\dots,0,\hat{\sigma}^a_1,\dots,\hat{\sigma}^a_{F_1-h_a})$, which we have diagonalised by a flavour rotation.
Then, setting to zero the VEV of the real scalar in the gauge vector multiplet for simplicity, the last $F_1-h_a$ of the fields  $(p_1)^\alpha$ and $(q_1)_\alpha$ in \eqref{W_flavoured_KW_vectorlike} acquire opposite real masses equal to $\pm\hat{\sigma}^a_1,\dots,\pm\hat{\sigma}^a_{F_1-h_a}$. Once similar real masses are given to the other flavours, in the way explained above, the superpotential and matter content of the resulting low energy theory is that of \eqref{W_flavoured_KW} and table above. The CS levels of the low energy theory, arising from integrating out the massive multiplets collectively denoted as $\psi$, depend on the signs of the spurionic VEVs, according to the shift 
\be
k_i \to k_i + \frac{1}{2} \sum\nolimits_\psi d_2(R_i[\psi])\, \mathrm{sign}(M_\psi) \;,
\ee
where $R_i[\psi]$ is the gauge representation under the $i$-th gauge group of a massive multiplet, and $d_2$ is twice the Dynkin index of the representation: $d_2(\square)=d_2(\overline{\square})=1$.
The procedure of giving real masses to flavours in a 3d KW theory with doubled vectorlike flavours and vanishing bare CS levels and integrating out massive multiplets to end up in a flavoured ABJM model ensures that the quantisation condition \eqref{level_quantisation} is fulfilled.

Real mass deformations coming from spurionic VEVs of real adjoints of the flavour symmetry groups are easily seen in the type IIB brane construction as certain deformations of the fivebrane webs of figures \ref{fig:chiral_flavours_x6} and \ref{fig:3d_KW_1_Ouyang_flav}. As we explained, the flavour groups $U(F_1)^2\times U(F_2)^2$ are supported by the semi-infinite D5 and $D5_\theta$ branes extending at $x_7<0$ and $x_7>0$. If we want to break $U(F_1)\to U(h_a)\times U(1)^{F_1-h_a}$ by real masses for $F_1-h_a$ of the fields  $(p_1)^\alpha$ and $(q_1)_\alpha$, for instance, it suffices to pull $F_1-h_a$ of the semi-infinite D5 extending at $x_7<0$ up or down, in such a way that they no longer intersect the D3 branes. 
As a consequence, $F_1-h_a$ pairs of $53$ and $3'5$ strings have a finite minimal length, yielding massive flavours. The sign of the real mass of a flavour field is positive or negative according to whether the string is oriented upwards or downwards in the $73$ plane (compare with the orientations of \eqref{fig:chiral_flavours_x6}). Therefore our convention  is that a half-D5 brane which is pulled down (up) corresponds to a positive (resp. negative) eigenvalue of $\langle \sigma_F \rangle$.

\begin{figure}[p]
\begin{center}
\includegraphics[width=12.5cm]{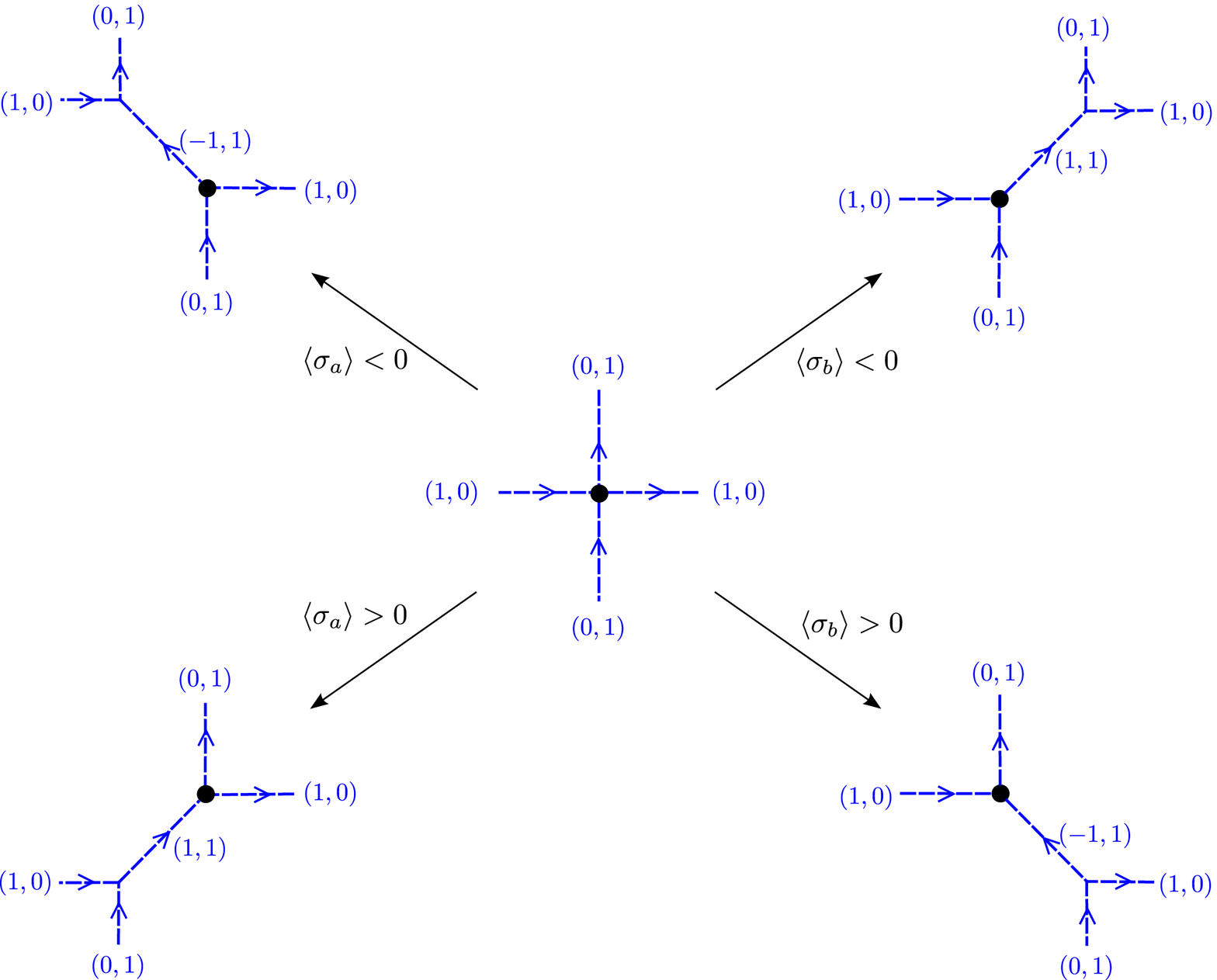}
\bigskip\bigskip

\includegraphics[width=12.5cm]{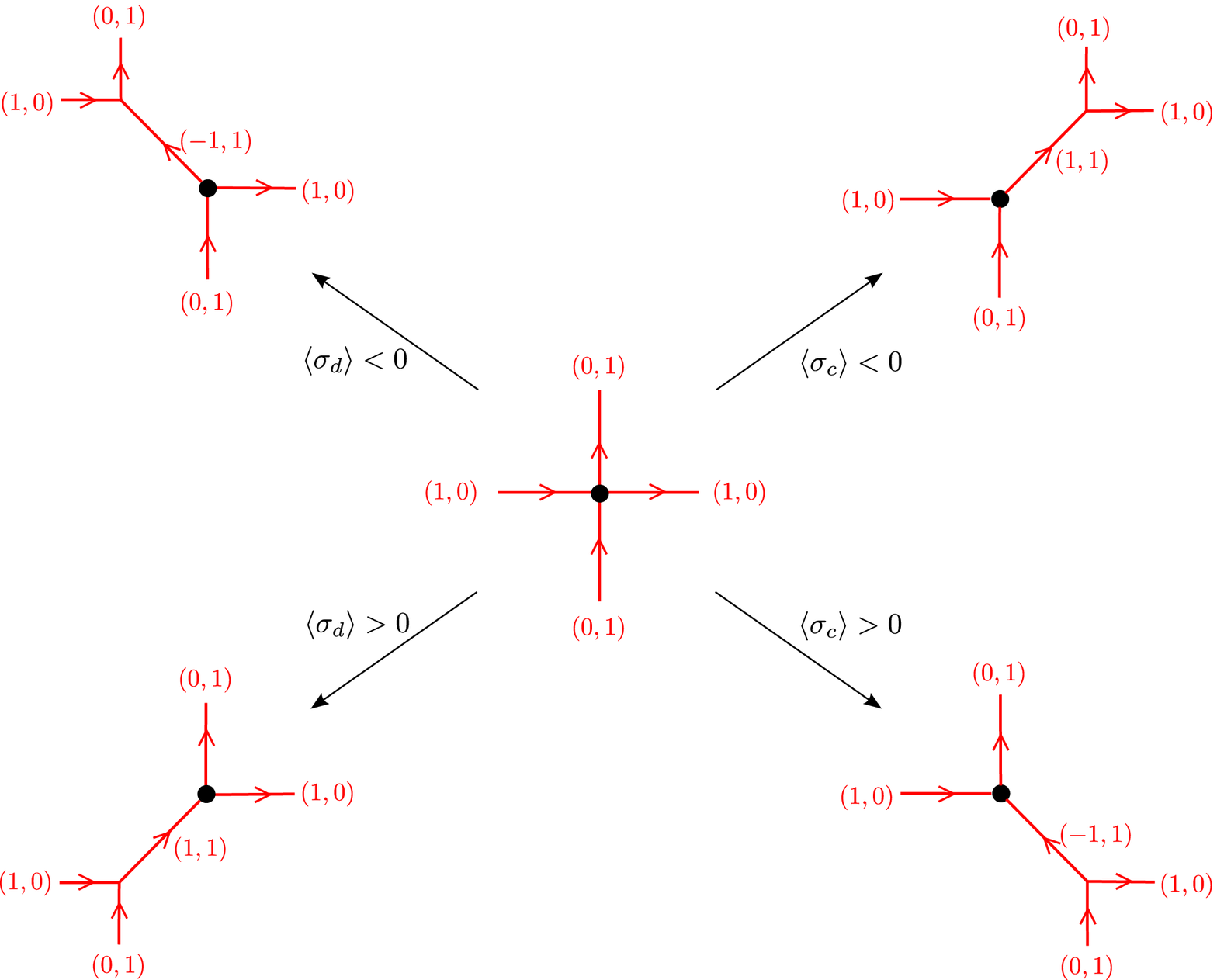}
\caption{\small Web deformations (in the 37 plane) of the NS5-D5 system that leave massless favours, and corresponding spurionic VEVs. D3 branes intersect the fivebrane webs at the black dot.}\label{fig:webs}
\end{center}
\end{figure}

Since such web deformations can be performed separately for each semi-infinite D5 brane, either pushing it up towards $x^3=+\infty$ or pulling it down towards $x_3=-\infty$, in drawing the web deformation we will focus on web deformation of an NS5-D5 intersection, with a single D5 brane divided in two halves by the NS5 brane. Further web deformations (once the additional flavour branes are reinstated) have analogous effects, as sketched above. The web deformations of the NS5-D5 systems that leave some of the flavours massless are shown in fig. \ref{fig:webs}, together with the spurionic VEVs $\langle \sigma_i\rangle$ ($i=a,b,c,d$) of the real scalar in the background vector multiplet of the flavour symmetry.

Comparing with figure \ref{fig:3d_KW_1_Ouyang_flav}, we can list the deformations of fig. \ref{fig:webs} and the effect on the 3d field theory as follows:
\begin{itemize}
\item $\langle \sigma_a\rangle <0$: masses $m_{p_1}<0$, $m_{q_1}>0$. \quad $\langle \sigma_a\rangle \to -\infty \Longrightarrow \delta \vec{k} \equiv (\delta k_1, \delta k_2)=(-\frac{1}{2},+\frac{1}{2})$.
\item  $\langle \sigma_a\rangle >0$: masses $m_{p_1}>0$, $m_{q_1}<0$. \quad $\langle \sigma_a\rangle \to +\infty \Longrightarrow \delta \vec{k} =(+\frac{1}{2},-\frac{1}{2})$.
\item $\langle \sigma_b\rangle <0$: masses $m_{\tilde p_1}<0$, $m_{\tilde q_1}>0$. \quad $\langle \sigma_b\rangle \to -\infty \Longrightarrow \delta \vec{k} =(+\frac{1}{2},-\frac{1}{2})$.
\item  $\langle \sigma_b\rangle >0$: masses $m_{\tilde p_1}>0$, $m_{\tilde q_1}<0$. \quad $\langle \sigma_b\rangle \to +\infty \Longrightarrow \delta \vec{k} =(-\frac{1}{2},+\frac{1}{2})$.
\item $\langle \sigma_c\rangle <0$: masses $m_{p_2}<0$, $m_{q_2}>0$. \quad $\langle \sigma_c\rangle \to -\infty \Longrightarrow \delta \vec{k} =(-\frac{1}{2},+\frac{1}{2})$.
\item  $\langle \sigma_c\rangle >0$: masses $m_{p_2}>0$, $m_{q_2}<0$. \quad $\langle \sigma_c\rangle \to +\infty \Longrightarrow \delta \vec{k} =(+\frac{1}{2},-\frac{1}{2})$.
\item $\langle \sigma_d\rangle <0$: masses $m_{\tilde p_2}<0$, $m_{\tilde q_2}>0$. \quad $\langle \sigma_d\rangle \to -\infty \Longrightarrow \delta \vec{k} =(+\frac{1}{2},-\frac{1}{2})$.
\item  $\langle \sigma_d\rangle >0$: masses $m_{\tilde p_2}>0$, $m_{\tilde q_2}<0$. \quad $\langle \sigma_d\rangle \to +\infty \Longrightarrow \delta \vec{k} =(-\frac{1}{2},+\frac{1}{2})$.
\end{itemize}

As a simple example of a web deformation, we can start with the configuration of section \ref{subsec:flavoured_KW} with $F_1=k$, $F_2=0$, and pull the $k$ semi-infinite D5 brane on the left down and push the $k$ semi-infinite D5 on the right up to infinity: we are thus left with a $(k,1)5$ brane and the NS5$_\theta$ brane.
The resulting quiver gauge theory is an $\calN=2$ version of ABJM (plus YM kinetic terms) with CS levels $(k,-k)$, with the quartic superpotential unrelated to the CS level. $\calN=3$ supersymmetry enhancement (leading to $\calN=6$ in the IR) is achieved by setting the angle $\theta$ in the 48 and 59 planes between the $(k,1)5$ brane and the NS5$_\theta$ brane equal to the angle $\arctan k$ in the 37 plane.%
\footnote{We assume here for simplicity that the axio-dilaton is at the self-dual value $\tau=i$. See \cite{Kitao:1999uj} for the generalisation.}
The actual value of $\theta\neq 0$ does not affect the algebraic description of the vacuum moduli space. This is the reason why we took the 3d KW theory rather than ABJM as a starting point. The two options are clearly equivalent for our purposes. 
Incidentally, notice that pulling a semi-infinite D5 brane in from, say, $x^3=+\infty$, and then pushing it out to $x^3=-\infty$ (or viceversa) is a way of converting an NS5 brane into a $(\pm 1,1)5$ brane, thus shifting the CS levels of the gauge theory by plus and minus one unit.

\begin{figure}
 \begin{minipage}[t]{7cm}
   \centering
\includegraphics[width=6cm]{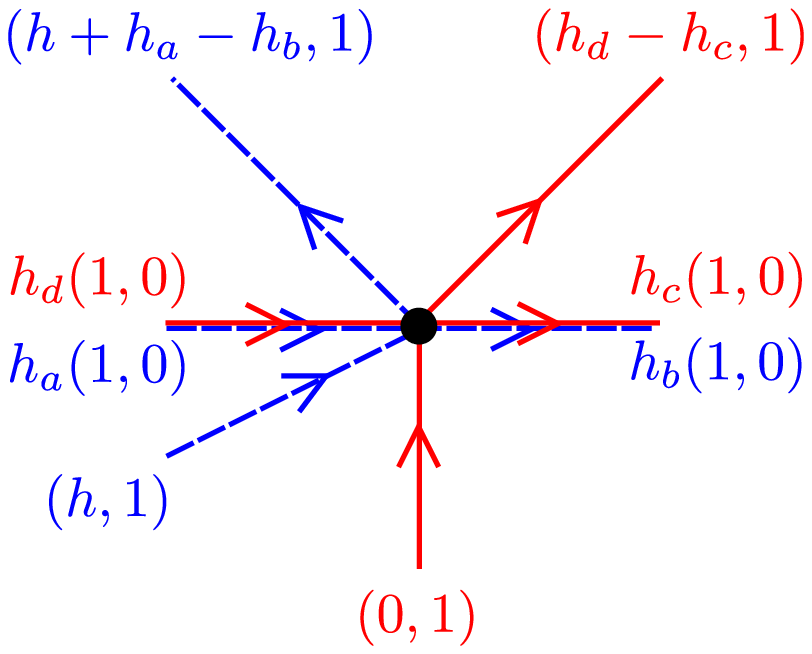}
\caption{\small 5-brane configurations in the 37 plane for flavoured ABJM. The plot is for $h=2$ and $(h_a,h_b,h_c,h_d)=(1,4,1,2)$. Suspended D3 branes are represented by the black dot intersecting the fivebrane junctions.}\label{fig:5brane_generic}
 \end{minipage}
 \ \hspace{2mm} \hspace{3mm} \
 \begin{minipage}[t]{7cm}
  \centering
\includegraphics[width=4cm]{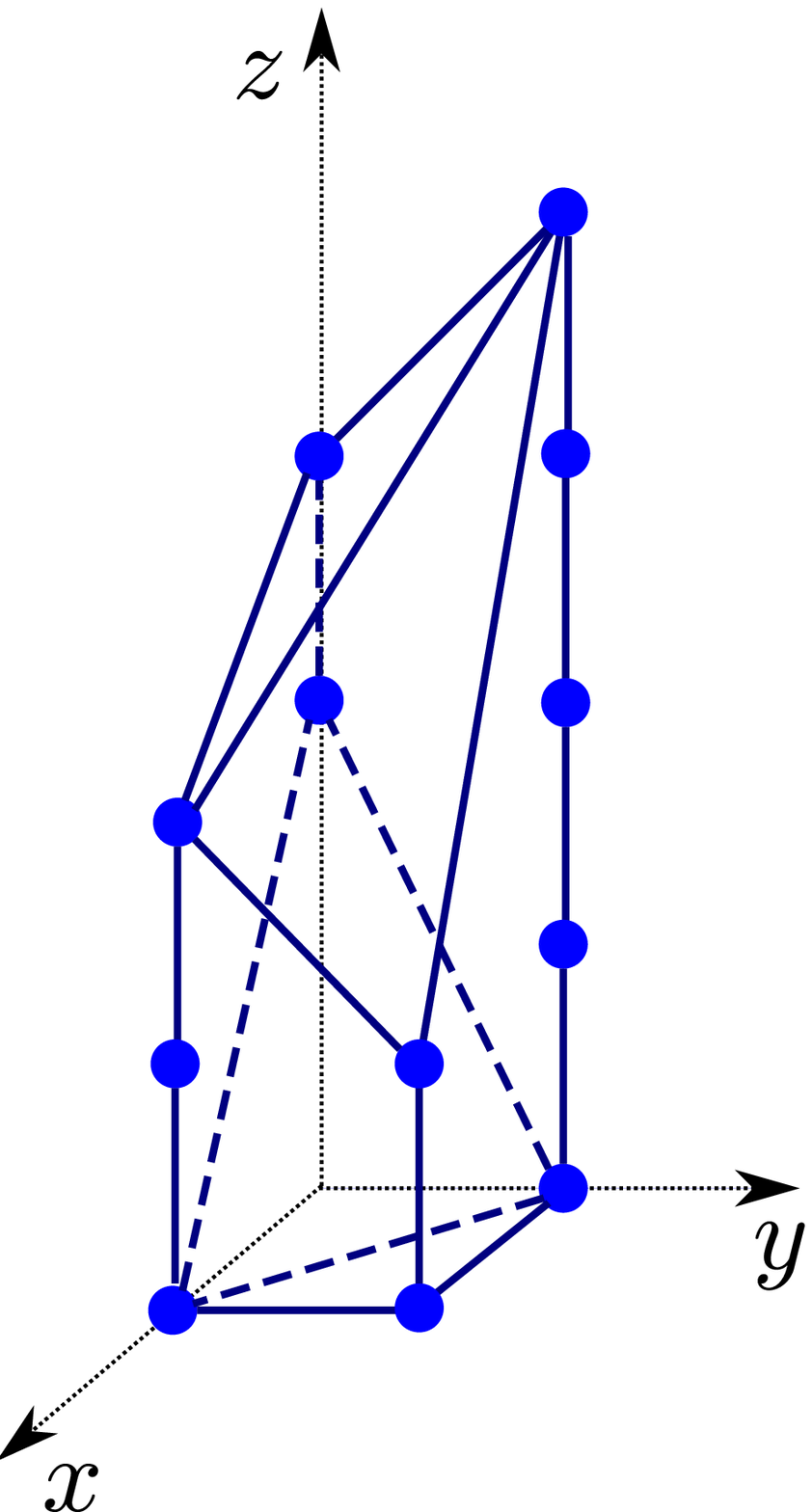}
\caption{\small Toric diagram of the conical Calabi-Yau fourfold dual to the fivebrane system of fig. \ref{fig:5brane_generic}.}\label{fig:toric_generic}
 \end{minipage}
\end{figure}

Thanks to the previous construction, we can finally introduce the type IIB brane configuration that engineers the flavoured ABJM theory with superpotential \eqref{W_flavoured_KW} and CS levels $(k,-k)$. It should be noted that given a field theory, the brane configuration is determined up to T-transformations in the S-duality group of type IIB, that leave D5 branes invariant. We fix this ambiguity with a particular choice that makes contact with \cite{Benini:2009qs}.
To this aim, let us introduce the following notation:
\begin{align}
N_L&\equiv h_a+h_c  \label{N_L}\\
N_R&\equiv h_b+h_d\\
h&\equiv k-\frac{1}{2}(N_L-N_R)\\
\hat{h}&\equiv k+\frac{1}{2}(N_L-N_R)=h+N_L-N_R\;, \label{hhat}
\end{align}
so that 
\begin{equation}\label{CS_level_from_h}
k=\frac{h+\hat{h}}{2}\;.
\end{equation}
Incidentally, note that in the abelian $U(1)_k\times U(1)_{-k}$ 3d gauge theory the charges of bifundamental fields and diagonal monopole operators $T$ and $\tilde T$ are \cite{Benini:2009qs}
\be\label{charges_flav_ABJM}
\begin{tabular}{c|c c c c}
                & $A_i$ & $B_j$ &  $T$  & $\tilde{T}$ \\ \hline
$U(1)_{\frac{h+\hat{h}}{2}}$    &  $1$  & $-1$  &  $\hat{h}$ &  $-h$    \\
$U(1)_{-\frac{h+\hat{h}}{2}}$ & $-1$  &  $1$  &  $-\hat{h}$ & $h$
\end{tabular}
\ee
and the quantum corrected F-term relation between monopole operators and bifundamentals reads 
\begin{equation}
\label{quantum_flav_ABJM}
T \tilde{T} = A_1^{h_a} B_1^{h_b} A_2^{h_c} B_2^{h_d} \;.
\end{equation}
The type IIB brane configuration can be taken to be as in fig. \ref{fig:5brane_generic}, as can be checked by suitably web-deforming the configuration of fig. \ref{fig:3d_KW_1_Ouyang_flav} and following the result in the 3d field theory.

These three-dimensional field theories flow to infrared fixed points at which the YM kinetic terms disappear. The geometric moduli spaces of (the abelian version of) such superconformal field theories (SCFTs) are conical toric Calabi-Yau (CY) fourfolds \cite{Benini:2009qs}, which are dual to the type IIB fivebrane system. More precisely, the type IIB system of two webs of fivebranes at an angle is dual to M-theory on the toric Calabi-Yau cone, and D3 branes probing the fivebrane system are dual to M2 branes probing the local Calabi-Yau. Given a torically flavoured ABJ(M) field theory specified by the previous data, the toric diagram of the toric $CY_4$ singularity which is the geometric moduli space of the SCFT has the following set of points in $\bbZ^3$:
\begin{equation}
\label{generic_toric_diag_for_flavored_ABJM}
a_{h+i} = (0,0,h+i) \;, \qquad  b_j= (0,1,j) \;, \qquad
c_l = (1,1,l) \;, \qquad d_m = (1,0,m) \;,
\end{equation}
where $i=0,\dots,h_a$, $j=0,\dots,h_b$, $l=0,\dots,h_c$, $m=0,\dots,h_d$.
See figure \ref{fig:toric_generic} for  the toric diagram of the Calabi-Yau fourfold dual to the fivebrane system of fig. \ref{fig:5brane_generic}. We will elaborate on the relation between toric diagram and fivebrane webs in section \ref{sec:webs-toric}.


\subsection{$\bbZ_2$ actions}\label{subsec:Z2}

Charge conjugation takes particles into antiparticles, and leaves CS terms invariant. If we combine it with the interchange of $A$ and $B$ (and of the corresponding flavours), namely if we consider the $\bbZ_2$ action
\begin{equation}
\calI\;:\qquad\qquad A_i \leftrightarrow B_i^\dagger\;,\qquad\qquad  p_i\leftrightarrow \tilde q_i^\dagger\;,\qquad\qquad q_i\leftrightarrow \tilde p_i^\dagger\;,
\end{equation}
we get a new flavoured ABJM field theory, with the same CS level $k$ and gauge group ranks, but the ranks of the flavour groups are interchanged: $(h_a,h_c)\leftrightarrow(h_b,h_d)$. Therefore $N_L\leftrightarrow N_R$, and $h\leftrightarrow \hat{h}$.
Clearly the geometric moduli space of the field theory does not change under this $\bbZ_2$ action, as can be easily verified.

On the other hand, parity (P) flips the sign of CS levels $k\leftrightarrow -k$, keeping flavour and colour groups fixed, therefore it acts on the field theory as $h\leftrightarrow -\hat{h}$, leaving $h_a$, $h_b$, $h_c$, $h_d$ fixed. It effectively interchanges the diagonal monopole operators, as implied by the pseudoscalar nature of the dual photon.
Again the geometric moduli space of the field theory does not change under this $\bbZ_2$ action.

In both cases a suitable $SL(4,\bbZ)$ transformation maps the toric diagram of the initial theory into the one of the final theory, with the $z$ axis reversed.


\subsection{Fivebrane webs, toric $CY_4$ and IIA configuration}\label{sec:webs-toric}

It is worth remarking how the fivebrane systems and the toric diagram of the toric Calabi-Yau fourfold are related. As well known, a web of $(p,q)$5 branes in type IIB string theory is dual to M-theory on a toric Calabi-Yau threefold whose toric diagram is the dual graph to the web diagram \cite{Aharony:1997bh,Leung:1997tw}.%
\footnote{See also \cite{Jensen:2009xh} for a recent discussion focused on dualities.}
A completely web-deformed $(p,q)$-web, such that all the junctions are between three fivebranes whose $(p,q)$ vectors are primitive, is dual to a complete triangulation of the toric diagram, corresponding to a crepant resolution of the toric $CY_3$ cone. Blowing down all the exceptional cycles corresponds to reducing the $(p,q)$ web to have a single junction, from which only semi-infinite fivebranes emanate.
Displacing subwebs at equilibrium transversely to the plane of the web is dual instead to a complex structure deformation of the $CY_3$.

The setup we are studying involves two fivebrane webs. Each one in isolation is dual to a toric $CY_3$. The whole fivebrane system is dual to M-theory on a toric $CY_4$ that was derived in \cite{Benini:2009qs} analysing the field theory (which in turn was motivated by reduction of M-theory on the fourfold to type IIA). The toric diagram of this fourfold can be sliced in two layers orthogonal to the $x$ axis. Each of the vertical layers (at $x=0$ and $x=1$ respectively) is the toric diagram of a toric $CY_3$ which is dual to one of the $(p,q)5$ brane webs.%
\footnote{These layers are vertical because the vertical direction in the ambient space of the toric diagram is related to the $U(1)$ isometry of the M-theory circle, in the mirror symmetry frame we are working in.}
Compare figures \ref{fig:5brane_generic} and \ref{fig:toric_generic} for an example.

Minimal web deformations that move a single semi-infinite D5 brane to $x^3=\pm\infty$, which are interpreted as infinite real mass terms for a single flavour pair in the 3d field theory, remove a single point in the toric diagram, either the one at the top or the one at the bottom of a vertical column. 

On the other hand, one could also consider a web deformation that moves a minimal subweb containing two semi-infinite $(p,1)5$ branes with different $p$ (together with the semi-infinite D5 branes that are joined to them by charge conservation) to $x^7=\pm\infty$. This is interpreted as an infinite FI term forcing a VEV for one or the other bifundamental (say $A_1$ or $B_1$) associated to the fivebrane in the 3d field theory. The VEV gives a complex mass to the flavours coupled to the bifundamental. An entire column is removed in the toric diagram, leaving the toric diagram of the geometric moduli space of $\calN=8$ SYM with a certain number of $\calN=2$ preserving massless flavours.

The type IIB brane cartoon also encodes the data of the type IIA background which arises after reducing M-theory along a circle action corresponding to the vertical direction in the toric diagram \cite{Aganagic:2009zk,Benini:2009qs}: the type IIA configuration is obtained from the type IIB configuration upon T-duality along the $x^6$ circle direction transverse to the fivebrane webs. It involves a (resolved) conifold fibred over a real line, with RR $F_2$ form flux though the $\bbC\bbP^1$, and $h_a$, $h_b$, $h_c$ and $h_d$ D6 branes embedded respectively along the four toric divisors of the singular conifold lying over the origin of the real line \cite{Benini:2009qs}. The real line in IIA corresponds to the $x^3$ coordinate in IIB. 
The K\"ahler parameter $\chi(x^3)$ of the 2-cycle of the conifold depends piecewise linearly on the coordinate $x^3$ of the real line. It can be read off in the IIB brane cartoon from the difference in the $x^7$ coordinates of the two fivebrane webs at a given $x^3$, see figure \ref{fig:5brane_generic}:
\begin{equation}\label{Kaehler_par}
\chi(x^3)=x^7_{blue}(x^3)-x^7_{red}(x^3)=\begin{cases}
\hat{h} x^3\;, & x^3>0\\
h x^3\;, & x^3<0
\end{cases}
\;.
\end{equation}
The RR 2-form flux through the 2-cycle is constrained by supersymmetry to equal the first derivative $\chi'(x^3)$
\begin{equation}
\frac{1}{2\pi}\int_{\mathbb{CP}^1(x^3)} F_2=\chi'(x^3)=\begin{cases}
\hat{h} \;, & x^3>0\\
h \;, & x^3<0
\end{cases}\;,
\end{equation}
and is therefore constant in $x^3$, with a jump at the location of the D6 branes $x^3=0$.

We end this section with some brief comments on the action of the S-duality group $SL(2,\bbZ)$ in type IIB and mirror symmetry of three-dimensional gauge theories. If the special linear matrix 
\begin{equation}
M=\begin{pmatrix}
a & b \\ c & d
\end{pmatrix} \in SL(2,\bbZ)
\end{equation}
acts on the fivebrane charge vectors as
\begin{equation}
\begin{pmatrix}
p\\q
\end{pmatrix}
\mapsto M \begin{pmatrix}
p\\q
\end{pmatrix}\;,
\end{equation}
then points in the 3d toric diagram are acted upon as
\begin{equation}
\begin{pmatrix}
x\\y\\z
\end{pmatrix}
\mapsto \begin{pmatrix}
1 & 0 & 0 \\
0 & d & -c\\
0 & -b & a
\end{pmatrix}
\begin{pmatrix}
x\\y\\z
\end{pmatrix}\;,
\end{equation}
which is an $SL(2,\bbZ)$ transformation in the $(y,z)$ plane. In particular, the effect on the toric diagram of the S-transformation can be visualised as a $\pi/2$ rotation about the $x$ axis.
The action on the three-dimensional field theories is to interchange mirror symmetric field theories. By these means, it is quite simple to find mirror symmetry pairs among simple flavoured ABJM models and flavoured versions of $\calN=8$ SYM. For instance, the gauge theory for M2 branes at the $D_3$ singularity can be equivalently described by the 3d KW model with a flavour pair coupled to one of the $A$ bifundamentals and a flavour pair coupled to one of the $B$ bifundamentals, or by $\calN=8$ SYM theory endowed with a single flavour pair coupled to each adjoint chiral superfield (explicitly breaking supersymmetry down to $\calN=2$) \cite{Benini:2009qs}. Similarly, the ABJM quiver theory with CS levels $\pm 1$ and with a flavour pair coupled to one of the $A$ bifundamentals and a flavour pair coupled to one of the $B$ bifundamentals is IR dual to the ABJM quiver theory with levels $\pm 1$ and one flavour pair per each of the $A$ bifundamentals. 

On the contrary, the T-transformation that leaves D5 branes invariant only modifies the RR charge $p$ of $(p,1)5$ branes, hence it acts trivially on the three-dimensional field theory.


\section{D3 brane creation and fractional M2 branes} \label{subsec:HWeffect_flavoured_ABJM}

In this section we move to the study of the brane creation phenomenon --- also known as Hanany-Witten (HW)  effect \cite{Hanany:1996ie} --- that occurs when the two fivebrane webs of section \ref{sec:flavoured_ABJ}  cross each other when sliding in the mutually transverse $x^6$ direction. The brane creation phenomenon is related to the ``$s$-rule'' that determines the maximal number of D3 branes that may be suspended between two fivebranes of different charge compatibly with supersymmetry. In elliptic models such as the ones under investigation, where the transverse $x^6$ direction is periodic, a modified $s$-rule applies, that follows from imposing the ordinary $s$-rule on the covering space of the circle \cite{Aharony:2009fc,Evslin:2009pk}.

This analysis, inspired by \cite{Aharony:2008gk,Aharony:2009fc,Evslin:2009pk}, allows us to determine the number of inequivalent SCFTs for a given quiver gauge theory with superpotential \eqref{W_flavoured_KW} (with fixed matter content, CS levels and minimal common gauge rank), and to draw some conclusions on duality cascades that lead to such supersymmetric IR fixed points.

\subsection{D3 brane creation}\label{subsec:brane_creation}

We begin the analysis with the creation of D3 branes when the two fivebrane systems of fig. \ref{fig:5brane_generic} cross each other as they slide along the mutually transverse $x^6$ direction. 

Recall that when two linked $(p_1,q_1)5$ and $(p_2,q_2)5$ branes cross, $|p_1 q_2 - p_2 q_1| $ D3 branes are created between them, oriented in the $x^6$ direction in the way that preserves supersymmetry, due to the Hanany-Witten effect \cite{Kitao:1998mf}. The situation we are interested in is more complicated as it involves fivebrane junctions. In order to avoid double counting, we resort to the trick of slightly displacing the junctions of the two fivebrane webs in the $37$ plane in such a way that there are only (possibly several) intersections of $(p_1,q_1)5$ and $(p_2,q_2)5$ branes.%
\footnote{In terms of the 3d field engineered by the brane system, the displacement between the two junction is interpreted as a FI term or a mass term for flavours.}
We then determine the total number $n$ of D3 branes that are created between the two fivebrane webs at the intersection points in the presence of this displacement, by applying the $s$-rule of \cite{Kitao:1998mf}.
We stress that this is only a trick. We are mostly interested in the configuration where there is a single intersection point at $x^3+ix^7=0$, which we can recover in the limit of vanishing deformation (see however section \ref{sec:Higgsing}). The limit is continuous, as the total number of D3 branes created at intersections is independent of the displacement vector. In the limit of vanishing displacement all the D3 branes are created at $x^3=x^4=x^5=x^7=x^8=x^9=0$, and span 012 as well as an interval between the two fivebrane systems along $x^6$.

With the fivebranes of fig. \ref{fig:5brane_generic}, the total number $n$ of D3 brane created depends on the 
relative orientation between the $(h,1)5$ and the $(0,1)5_\theta$ branes extending from $x^3=-\infty$ and between the $(h+h_a-h_b,1)5$ and the $(h_d-h_c,1)5_\theta$ branes extending to $x^3=+\infty$, see fig. \ref{fig:Hanany-Witten}. Those relative orientations are determined by the signs of $h$ and $\hat{h}\equiv h+h_a-h_b+h_c-h_d$, and the result for the total number of D3 branes created by HW effect as the fivebranes cross each other is: 
\begin{figure}[t]
\begin{center}
\includegraphics[width=14cm]{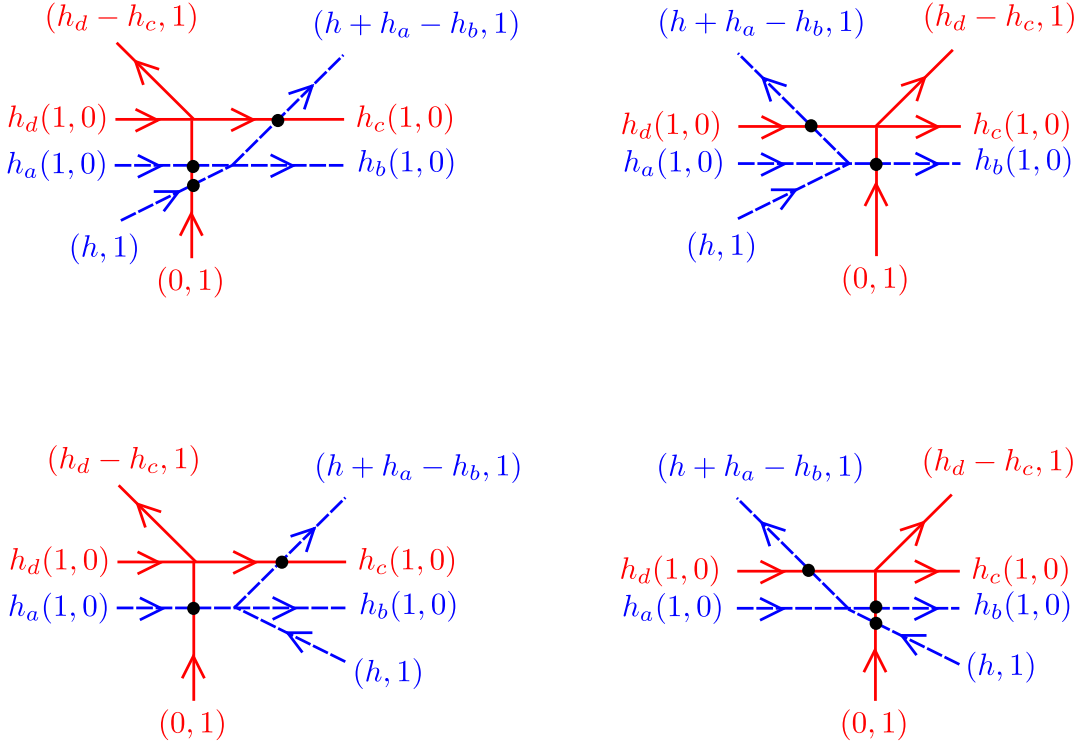}
\caption{\small Fivebrane configurations and brane creation effect for flavoured ABJM models: stacks of suspended D3 branes are created at the black dots when the two fivebrane systems cross each other.}\label{fig:Hanany-Witten}
\end{center}
\end{figure}
\begin{enumerate}
\item $h\geq 0$, $\hat{h}\geq 0$: \quad  $n=h+h_a+h_c=\hat{h}+h_b+h_d$ D3 branes are created; 
\item $h\geq 0$, $\hat{h}\leq 0$: \quad  $n=h_b+h_d$ D3 branes are created;
\item $h\leq 0$, $\hat{h}\geq 0$: \quad  $n=h_a+h_c$ D3 branes are created;
\item $h\leq 0$, $\hat{h}\leq 0$: \quad  $n=-h+h_b+h_d=-\hat{h}+h_a+h_c$ D3 branes are created.
\end{enumerate}
Using the notation of formulae \eqref{N_L}-\eqref{hhat}, we rewrite these four cases respectively as 
\begin{enumerate}
\item $k\geq +\frac{1}{2}\,|N_L-N_R|$: \quad  $n=k+\frac{1}{2}(N_L+N_R)$ D3 branes are created; 
\item $|k|\leq \frac{1}{2}\,(N_R-N_L)$: \quad  $n=N_R$ D3 branes are created;
\item $|k|\leq \frac{1}{2}\,(N_L-N_R)$: \quad  $n=N_L$ D3 branes are created;
\item $k\leq -\frac{1}{2}\,|N_L-N_R|$: \quad  $n=-k+\frac{1}{2}(N_L+N_R)$ D3 branes are created.
\end{enumerate}
The latter notation is useful if one is interested in engineering 3d $\calN=2$ $U(N_c)$ CS-SQCD with fundamentals and antifundamentals in arbitrary numbers, by stretching $N_c$ D3 branes between the two fivebrane webs along an interval of an $x^6$ line.

Note that the $\bbZ_2$ actions of section \ref{subsec:Z2} act as follows on the previous classes: 
\begin{equation}
P: \quad (1,2,3,4)\mapsto (4,2,3,1)\;,\qquad\qquad \calI: \quad (1,2,3,4)\mapsto (1,3,2,4)\;.
\end{equation}

\subsection{Fractional M2 branes and inequivalent SCFTs}\label{subsec:fracM2}

Choosing a fivebrane configuration among those in fig. \ref{fig:5brane_generic} amounts to selecting a specific three-dimensional quiver gauge theory with superpotential \eqref{W_flavoured_KW}, determined by the CS levels and the ranks of the flavour group. The number of D3 branes $N_1$ and $N_2$ suspended along the two intervals between the fivebranes specifies the ranks of the gauge groups, so that the 3d field theory is a $U(N_1)_k\times U(N_2)_{-k}$ quiver gauge theory.

When $N_1=N_2=N$, namely when there are $N$ D3 branes wrapping the circle and no D3 branes on an interval, the field theory flows to an $\calN=2$ supersymmetric IR fixed point, which describes the low energy physics on a stack of $N$ M2 branes at a toric conical Calabi-Yau fourfold $C(Y_7)$ arising as the geometric moduli space of the conformal field theory. Such a fixed point has a large $N$ Freund-Rubin holographic dual in M-theory of the form $AdS_4\times Y_7$, where $Y_7$ is a toric Sasaki-Einstein 7-manifold (or orbifold). The rank $N$ in the field theory translates to the 7-form flux $\star_{11} G$ on $Y_7$ in M-theory. The M-theory background is the near horizon limit of $N$ M2 branes at the $CY_4$ singularity.

The discussion so far has paralleled what happens in the $AdS_5/CFT_4$ correspondence, where a fixed point is reached only if the ranks of all the gauge groups in the quiver are equal, corresponding to having only regular D3 branes probing a conical Calabi-Yau threefold in type IIB string theory. A novelty of the $AdS_4/CFT_3$ correspondence, however, is that $H^4(Y_7,\bbZ)$ may have nontrivial torsion, as opposed to simply connected toric Sasaki-Einstein 5-manifolds, whose cohomologies have no torsion \cite{Benishti:2009ky}. If a flat torsion $G$-flux is turned on, the BPS equations for vanishing supersymmetry variations and the Bianchi identities, which together imply the equations of motion, are still satisfied. 
If the order of $H^4_{tor}(Y_7,\bbZ)$ is $n$, then there are $n$ distinct $AdS_4\times Y_7$ M-theory backgrounds with the same non-torsion flux but different torsion $G$-flux \cite{Aharony:2008gk}: each unit of torsion $G$-flux is due to the backreaction of a fractional M2 brane which is an M5 branes wrapped on a torsion 3-cycle.
These backgrounds are dual to $n$ distinct conformal field theories, which in the classic example of \cite{Aharony:2008gk} are characterised by different ranks of the gauge group, with the same (common) minimal rank equal to the fixed number $N$ of regular M2 branes. The existence of a dual type IIB brane construction allowed the authors of \cite{Aharony:2008gk} (ABJ) to compute the number $n$ of inequivalent SCFTs with the same geometric moduli space: it was given by the number of D3 branes created in a Hanany-Witten transition in the type IIB engineering. 

A similar reasoning carries over into the flavoured ABJ models studied in this paper. The number $n$ of inequivalent SCFTs with given flavoured ABJ quiver, CS levels, superpotential \eqref{W_flavoured_KW} and common gauge rank is expected to equal the number of D3 branes which are created in a Hanany-Witten transition in the type IIB engineering, that we computed in section \ref{subsec:brane_creation} \cite{Aharony:2008gk}. The explanation is identical to the one presented in \cite{Aharony:2008gk}, and later refined in \cite{Aharony:2009fc,Evslin:2009pk}, except that here the number of D3 branes created in a HW transition does not necessarily equal the absolute value of the CS level. 
We refer the readers to those references and to the recent article \cite{Hashimoto:2010bq} which contains a nice review, and content ourselves with making some remarks and then stating the results of such an analysis.

The geometric moduli space $C(Y_7)$ of the 3d flavoured $\calN=2$ SCFT (and therefore the dual $AdS_4\times Y_7$ geometry) is determined by the type IIB fivebrane configuration, which also determines the matter content of the flavoured ABJ quiver, the CS levels and the superpotential \eqref{W_flavoured_KW}, up to a possible subtlety that we discuss below. The different superconformal field theories defined by those data are distinguished by the ranks of the gauge groups, corresponding to the number of D3 branes suspended between the fivebrane webs along the two arcs of the type IIB circle. D3 branes wrapping the circle are dual to regular D2 branes in IIA, which lift to regular M2 branes in M-theory. D3 branes extending along an interval only are dual to fractional D2 branes in IIA, namely D4 branes wrapping the vanishing $\bbC\bbP^1$ of the conifold (or wrapped anti-D4 branes with one unit of worldvolume gauge flux), and lift to M5-branes wrapping a 3-cycle in M-theory. The wrapped 3-cycle must be a torsion cycle if the field theory is superconformal and has an $AdS_4$ dual, otherwise the M5 brane would source a $G$-flux which spoils the $AdS_4$ geometry. 

Given a fivebrane configuration as in fig. \ref{fig:5brane_generic}, the number of D3 branes created in a HW transition $n$ depends on the CS level and the ranks of the flavour groups in the way computed in the previous subsection. As we wrote, this same $n$ is the number of inequivalent SCFTs for a given quiver gauge theory with fixed CS levels $\pm k$, number of flavours $\{h_F\}$ ($F=a,b,c,d$), minimal gauge rank $N$ and superpotential \ref{W_flavoured_KW}. More precisely, the SCFTs have gauge sector $U(N+M)_k\times U(N)_{-k}$, with $M=0,1,\dots,n$, and are infrared dual to theories with gauge sector $U(N+n-M)_{-k}\times U(N)_k$, so that $M=n$ is equivalent to $M=0$. For later convenience, we denote such theories as lying in the \emph{conformal window}, which is $|N_1-N_2|\leq n$. This is nothing but the outcome of the naive $s$-rule that does not take into account the fact that the $x^6$ direction is a circle \cite{Aharony:2008gk}. Note that the previous equivalence relates quivers in the conformal window, with the same minimal rank $N_{min}=N$.


In light of the results on the conformal window found in the type IIB brane construction and the remarks made above regarding the duality to M-theory and torsion $G$-fluxes, we expect that $H^4_{tor}(Y_7,\bbZ)\supseteq \bbZ_n$, where $Y_7$ is the Sasaki-Einstein link of the toric $CY_4$ cone dual to the fivebrane system, and $n$ the number of D3 branes created in the Hanany-Witten effect that we calculated in the previous section. We wrote $\supseteq$ rather than $=$ because we are considering the possibility that we are missing some SCFTs dual to $AdS_4\times Y_7$ geometries.  

On the gravity side, the case of singular $Y_7$ is subtle because supergravity must be supplemented with additional fields, and purely geometric and topological considerations may fail. On the field theory side, when both $A$ and $B$ fields are coupled to flavours, it is possible to switch on complex masses for some of the fundamental flavours, leading to a different low energy field theory with flavours coupled to composite bifundamentals such as $AB$ in the superpotential, and a smaller flavour symmetry. The geometric moduli spaces of the resulting field theories appear to coincide with those of the theories before mass deformation \cite{Benini:2009qs}, although it is not entirely clear in field theory that the mass deformed theories have a fixed point. The possibility of adding complex masses for vectorlike flavours occurs when the $CY_4$ cone has non-isolated singularities, and corresponds to turning on higher dimensional fields living at the non-isolated singularity.
In view of to these issues, we cannot exclude that when $Y_7$ is singular the number of field theories dual to $AdS_4\times Y_7$ backgrounds of given curvature is larger than the number of superconformal field theories that we found in type IIB in the context of the class defined by \eqref{W_flavoured_KW}. We will address further issues arising when $Y_7$ is singular in section \ref{subsec:frac_branches}.

On the other hand, if $Y_7$ is smooth the issues mentioned above disappear. In particular, the field theory appears to be uniquely determined by the fivebrane embedding in IIB, or equivalently by the IIA background with D6 branes.
It seems then reasonable to conjecture that in such cases the type IIB brane analysis captures all the field theories dual to $AdS_4\times Y_7$ backgrounds of given curvature, differing by torsion $G$-fluxes: if so, $H^4_{tor}(Y_7,\bbZ)= \bbZ_n$, where $n$ is the number of D3 branes created in a HW transition in type IIB. In section \ref{sec:smooth} we will present two one-parameter families of smooth $Y_7$ geometries arising in our construction, and we will check our conjecture in one of them containing $Y^{1,2}(\mathbb{CP}^2)$, as well as in $Q^{1,1,1}$.


\subsection{Duality cascades}\label{subsec:cascades}

If the gauge ranks in the UV $N_1^{UV}$ and $N_2^{UV}$ lie outside the conformal window, the fate of the theory depends on whether or not the modified $s$-rule of \cite{Aharony:2009fc,Evslin:2009pk} is satisfied. The modified $s$-rule is just the ordinary $s$-rule applied to the covering space of the circle, so that at most $n$ D3 branes can be attached to a fivebrane and any image of the other fivebrane in the covering space. Let us assume that in the UV $N_1^{UV}>N_2^{UV}$, which can always be achieved by choosing an appropriate parity frame, and define $N^{UV}=N_2^{UV}$ and $M^{UV}=N_1^{UV}-N_2^{UV}$, so that we start at high energy with a quiver whose gauge sector is $U(N^{UV}+M^{UV})_k\times U(N^{UV})_{-k}$.
The modified $s$-rule then reads 
\begin{equation}\label{modified_s_rule}
2n N^{UV}\geq M^{UV}(M^{UV}-n)\;.
\end{equation} 
If it is satisfied, it is natural to conjecture as in \cite{Aharony:2009fc,Evslin:2009pk} that the UV field theory flows to one of the SCFTs in the conformal window by a cascade of Seiberg dualities, which decrease the minimal common gauge rank $N$ and the difference of ranks $M$. 

If the flavour content of the quiver gauge theory is vectorlike, this Seiberg duality is the one of \cite{Giveon:2008zn} or \cite{Aharony:1997gp}. If instead the number of fundamentals does not equal the number of bifundamentals, such 3d Seiberg dualities need to be generalised.  Study of the HW brane transition between two fivebrane webs as in fig. \ref{fig:5brane_generic} suggests that the electric 3d $\calN=2$ $U(N_c)_k$ YMCS SQCD with $N_R$ fundamentals and $N_L$ antifundamentals should be dual to a magnetic $U(n-N_c)_{-k}$ SQCD with $N_R$ antifundamentals and $N_L$ fundamentals, likely together with dual meson singlets and a magnetic superpotential, and with $n$ as computed in the previous subsection. It is harder to read the magnetic singlets and superpotential off the type IIB construction with fivebrane webs, although we expect them to appear (whenever electric mesons exist) for consistency with known 3d Seiberg dualities. The study of this more general 3d Seiberg duality, which can be carried out by following the effect of real mass deformations on the aforementioned known dual pairs, is under current investigation and will be presented separately \cite{Cremonesi:2010}.

For now, suffice it to focus on its effects on the ranks of the gauge groups, which is easy to extract from the brane construction. Under $m$ such Seiberg dualities towards the IR, 
\begin{equation}\label{single_Seiberg_duality}
(N,\,M)\longrightarrow (N-m M+\frac{m(m+1)}{2}\,n,\,M-mn)\;.
\end{equation}
As in \cite{Aharony:2009fc,Evslin:2009pk}, the modified $s$-rule \eqref{modified_s_rule} is invariant under Seiberg dualities \eqref{single_Seiberg_duality}, which are realised as dynamical HW transitions in the type IIB engineering. The excess of D3 branes on one interval pulls the fivebranes thereby reducing $b$. Every time $b$ reaches an integral value, a dynamical HW transition (Seiberg duality) occurs. The cascade continues until the conformal window is reached, after which the distance $b$ between the fivebranes asymptotes a definite value. 

In the type IIA T-dual frame, the same RG flow should be described by a supergravity solution for (decoupled) regular D2 branes and fractional D2 branes  on the conifold fibred over a transverse line, with $h_a$, $h_b$, $h_c$ and $h_d$ D6 branes along each of the four toric divisors of the conifold respectively. The fractional D2 branes are D4 branes wrapped on the vanishing 2-cycle of the singular conifold.
In the absence of fractional D2 branes, the holographic RG flow solution develops a large effective string coupling in the IR. In that region the solution is best described in M-theory where the M-theory circle opens up, and asymptotes $AdS_4\times Y_7$ without torsion flux. In the presence of a number of fractional D2 branes such that the naive $s$-rule is not satisfied, the period of the $B$-field on the exceptional $\bbP^1$ of the conifold 
\begin{equation}
b=\frac{1}{4\pi^2 \alpha'}\int_{\bbP^1}B_2
\end{equation}
runs and crosses integer values at which fractional D2 branes become tensionless, corresponding to infinite YM coupling for one of the gauge groups in the dual field theory, after which one needs to resort to a Seiberg dual description.
 The transition corresponds to a decrease in the number of both regular and fractional D2 branes, as in \eqref{single_Seiberg_duality}. 
Still, the solution for the holographic RG flow is expected to have an M-theory lift which asymptotes $AdS_4\times Y_7$ with torsion flux (unless the RHS of \eqref{modified_s_rule} vanishes): there are as many units of torsion $G$-flux on $Y_7$ as the number of fractional M2 branes which survive in the IR modulo $n$. We stress that the fractional M2 branes that we refer to here are M5 branes wrapped on torsion 3-cycles, not on any supersymmetric 3-cycles.

If instead the gauge ranks violate the modified $s$-rule, the RG flow may still involve a duality cascade, but a nontrivial IR fixed point is not reached. 
The sequence of Seiberg dualities leads to $N<0$ before reaching $0\leq M\leq N$. Supersymmetry is expected to be broken when this happens. 

We should mention at this point that, at least in the quiver gauge theories with vectorlike flavours, for which the effect of 3d Seiberg duality is well established, due to the superpotential \eqref{W_flavoured_KW_vectorlike} some of the singlets which are generated under Seiberg dualities do not acquire a mass, and remain as light degrees of freedom, coupled to the magnetic flavours via superpotential terms.%
\footnote{As in 4d, after $m$ Seiberg dualities there are order $m$ singlets. However, all but the singlets which resulted from the last Seiberg duality are only coupled via irrelevant superpotential terms, and therefore can be ignored if one is interested in the IR limit of the RG flow as we are.}
See \cite{Benini:2007kg} for an analysis of this phenomenon, originally performed in the context of a 4d flavoured Klebanov-Strassler theory. 
The trilinear superpotential coupling between each of the singlets appearing after the last Seiberg duality and flavours can potentially be marginal at the IR fixed point. If so, the full superpotential at the fixed point would depend on whether or not the gauge theory has undergone Seiberg duality during the RG flow. It would be interesting to investigate this peculiarity in more depth, and understand whether it increases the number of inequivalent superconformal field theories. 

We remark that if the quiver gauge theory has vectorlike flavours, the geometric moduli space is a Calabi-Yau fourfold with non-isolated singularity, and the generation of singlets is related to this fact. A detailed discussion of this aspect from the point of view of intersecting flavour branes can be found in \cite{Benini:2007kg}.
Conversely, absence of non-isolated singularities requires the dual quiver gauge theory to have only fundamentals (or only antifundamentals) for each of the gauge groups, as will be explained in section \ref{sec:smooth}. In such a situation, no dual singlets can appear in 3d Seiberg duality and the subtlety is avoided.

\subsection{Caveat: fractional brane branches}\label{subsec:frac_branches}

In some of our flavoured model there is an important caveat to the conclusion that the RG flow ends up either in a nontrivial SCFT or in supersymmetry breaking. This applies whenever D3 branes may be suspended between two fivebrane segments with the same charges and (locally) with the same embedding in the 37 plane. A necessary condition for this to happen is that the Calabi-Yau fourfold in M-theory has a non-isolated singularity.
In such a situation, the $s$-rule does not apply \cite{Bergman:1999na}: classically, any number of D3 branes can be suspended between two fivebranes with the same charges. However, an instanton generated superpotential \cite{Affleck:1982as} generically lifts this classical Coulomb (or magnetic Coulomb) branch parametrised by the motion of fractional branes along the non-isolated singularity. Suppose for definiteness that in the type IIB brane picture the two fivebranes are NS5; then the elementary instantons, which are monopole configurations, are Euclidean D1 branes stretched between two adjacent D3 branes and the NS5 brane \cite{deBoer:1996ck}. The vector multiplet provides two fermionic zero modes in the background of an elementary instanton/monopole, and the monopole contributes a runaway term to the superpotential, lifting the whole classical Coulomb branch of fractional D-branes. If instead the Euclidean D-string intersects a flavour semi-infinite D5 brane there is no contribution to the superpotential, because the flavour fields furnish additional fermionic zero modes; then a moduli space for fractional branes can survive, even after taking into account all the nonperturbative corrections to the superpotential \cite{deBoer:1997kr,Tong:2000ky}. This requires that a number of D3 branes compatible with the (modified) $s$-rule stretches between a fivebrane junction and the other fivebrane (possibly with a junction too), and that at most one%
\footnote{At most one in order for the monopoles not to lift the branch.} 
D3 brane is suspended between any two segments of equally charged fivebranes. 

These potential fractional brane branches deserve further thought. 
It is an interesting question to ask what happens in the limiting case of vanishing VEV in one such branch, provided it exist. We are considering the case where a mobile suspended D3 brane reaches the stack of the other D3 branes which are suspended between two junctions or a junction and a fivebrane. Is the final configuration still restricted by the $s$-rule for fivebrane junctions that we derived for fivebrane junctions as a limit of the $s$-rule for fivebranes, or not? We are pondering this possibility because we are worried that the analysis of sections \ref{subsec:brane_creation} and \ref{subsec:fracM2} may be too naive when a D3 brane stretched along an interval can leave a junction and move along along fivebranes of the same charges which are parallel in 37.

If the final configuration  does not break the $s$-rule for D3 branes ending on fivebrane junctions that we used in the previous section, we are confident that it corresponds to a conformal field theory. 
If however the initial configuration saturated that $s$-rule, then the possibility of having one more D3 brane, violating the bound by one unit,%
\footnote{For later convenience we call this situation a \emph{minimal violation} of the $s$-rule.}
yet yielding a supersymmetric conformal field theory, would be puzzling: it would seem, by iterating the same argument, that in this situation there is no upper bound on the number of inequivalent SCFTs.  It is not hard to convince oneself that this possibility is not tenable, by checking against an example where the dual field theory was carefully studied. The example that we employ to this aim is a type IIB brane engineering of 3d $\calN=2$ $U(N_c)$ SQCD with $N_f$ flavours by means of $N_c$ D3 branes stretched between two identical fivebrane webs (at an angle) made of  a single junction of an NS5, $N_f$ D5 and an $(N_f,1)5$, all semi-infinite. The naive $s$-rule for fivebrane junctions that we derived as a limit of the $s$-rule for fivebranes precisely accounts for the conformal window $N_c\leq N_f$ \cite{Aharony:1997bx,Aharony:1997gp}. 

The phase structure of 3d $\calN=2$ SQCD also suggests a way out to the puzzle that the limit of a supersymmetric branch could be not supersymmetric. When $N_f=N_c-1$, the vacuum moduli space, which classically consists of intersecting Coulomb and mesonic branches, is smoothed out by a quantum deformation. As a result, supersymmetry is not broken, but the field theory does not reach a nontrivial fixed point. Notice that in terms of the brane construction, this is precisely a situation of minimal violation of the $s$-rule: perturbatively, $N_f=N_c-1$ D3 branes are allowed to be suspended between the two junctions, and another D3 to be mobile elsewhere. When $N_f<N_c-1$ supersymmetry is instead broken.
We expect a similar smoothing of fractional brane moduli spaces to happen more generally when there is $\calN=2$ supersymmetry and a minimal violation of the $s$-rule.

One might wonder in fact that the limit of vanishing VEV considered above does not actually exist in cases that marginally violate the $s$-rule: this would happen if in such cases the fractional brane branch of the moduli space is deformed at the quantum level, so that it no longer intersects the origin of the geometric branch, which on the other hand is not deformed. A somewhat similar phenomenon to the one that we are envisaging occurs for fractional D3 branes of $\calN=2$ type (which are D5 branes wrapping a vanishing non-rigid 2-cycle, with a one-complex-dimensional moduli space) in the $AdS_5/CFT_4$ correspondence: the Calabi-Yau threefold is not deformed, but the fractional branes source and transmute into a twisted sector flux along the non-isolated singularity, which does not allow them to coincide \cite{Bertolini:2000dk,Polchinski:2000mx,Benini:2008ir,Cremonesi:2009hq}.
Here we are also discussing fractional branes which are mobile at non-isolated singularities, and we are imagining that the phenomenon that occurs for $\calN=2$ fractional D3-branes in type IIB happens as well in M-theory  for M5 branes wrapped on vanishing 3-cycles (mobile along non-isolated singularity), when they exceed by one unit the number of torsion 3-cycles. 
The situation in M-theory is much less clear than in type II string theory, and it is hard to go beyond these qualitative analogies. 
It would be very interesting to pursue these thoughts further, perhaps studying the field theory side which is under better control.


\subsection{Fractional M2 branes and torsion $G$-flux: smooth $SE_7$ geometries}\label{sec:smooth}

We conclude this section applying the analysis of subsections \ref{subsec:brane_creation} and \ref{subsec:fracM2} to smooth Sasaki-Einstein 7-dimensional geometries.
Among the toric $CY_4$ cones which are the geometric moduli spaces of flavoured ABJ(M) theories that we have studied, some only have an isolated singularity, so that the link of the cone is a smooth Sasaki-Einstein 7-fold. 
A necessary and sufficient condition for the absence of non-isolated singularities is that all faces of the toric diagram are triangles, and the toric diagram contains no points inside faces or edges --- namely, all the integer points are vertices of the polyhedron --- \cite{Benishti:2009ky}.

\begin{figure}
 \begin{minipage}[t]{7cm}
   \centering
\includegraphics[width=4cm]{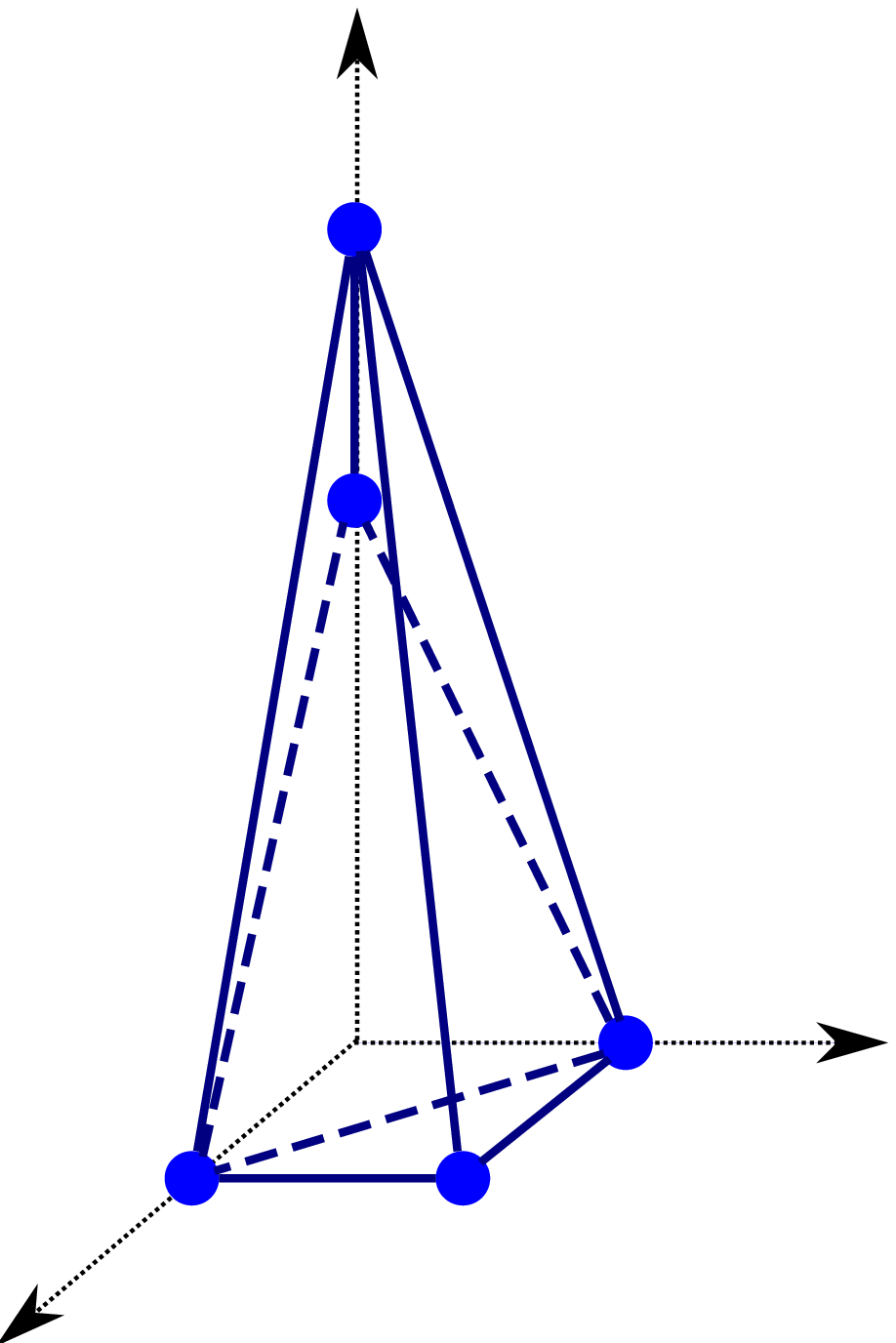}
\caption{\small Toric diagram for $L^{1,3,1,1}$.}\label{fig:toric_L1311}
 \end{minipage}
 \ \hspace{2mm} \hspace{3mm} \
 \begin{minipage}[t]{7cm}
  \centering
\includegraphics[width=4cm]{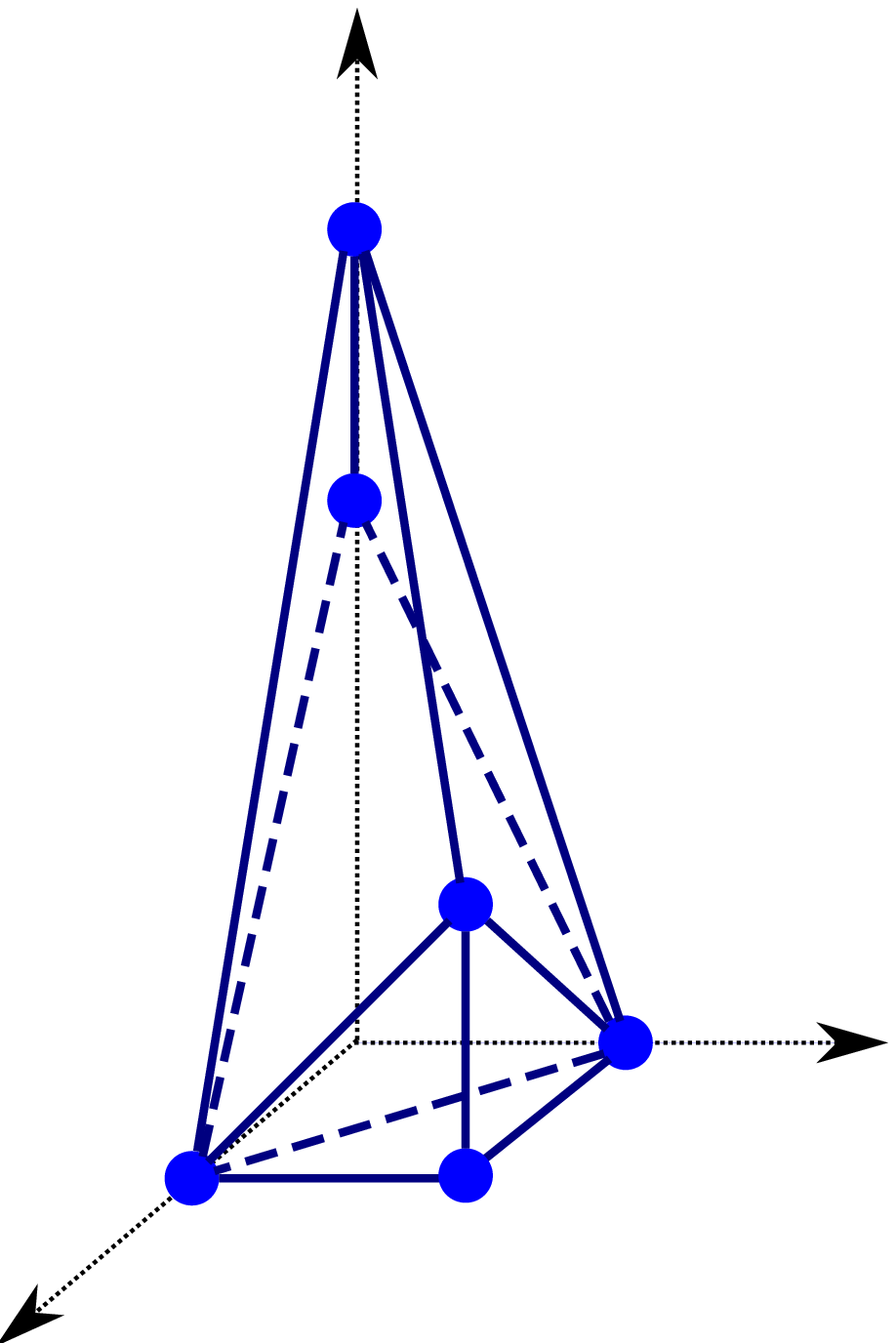}
\caption{\small Toric diagram for $Y^2$.}\label{fig:toric_family2_heq2}
 \end{minipage}
\end{figure}

Among the 3-dimensional toric diagrams that project to the toric diagram of the conifold, the class that we are studying, there are two one-parameter families that satisfy the previous requirement.
The first one has the toric diagram
\begin{equation}\label{toric_L_family}
 a_h = (0,0,h),\quad a_{h+1}=(0,0,h+1),\quad b_0=(0,1,0),\quad c_0=(1,1,0),\quad d_0=(0,1,0),
\end{equation}
where we can take $h\geq 1$ without loss of generality, see fig. \ref{fig:toric_L1311}.
The corresponding GLSM has charges 
\be\label{GLSM_L_family}
\begin{tabular}{c c c c c}
$c_0$  & $a_h$ &   $b_0$ & $d_0$ & $a_{h+1}$\\ \hline
 $-1$ &  $-(h+1)$ &$1$ & $1$ & $h$
\end{tabular}
\ee
The links of the cones are smooth Sasaki-Einstein 7-folds which span a one parameter subfamily $L^{1,h+1,1,1}$ of a four parameter family $L^{p,q,r_1,r_2}$ of smooth toric Sasaki-Einstein 7-folds whose metrics were found in \cite{Cvetic:2005ft}. The GLSM for the 4-parameter family of toric Calabi-Yau fourfolds with isolated singularities has charges $(-p,-q,r_1,r_2,p+q-r_1-r_2)$.
Note that when $h=1$, the Sasaki-Einstein reduces to $L^{1,2,1,1}=Y^{1,2}(\mathbb{CP}^2)$.

The dual gauge theories have ABJ(M) quivers, with CS levels $(h+\frac{1}{2},-(h+\frac{1}{2}))$, a single flavour pair coupled to the bifundamental $A_1$, and superpotential \eqref{W_flavoured_KW} with $h_a=1$ and $h_b=h_c=h_d=0$. According to the discussion of section \ref{sec:flavoured_ABJ}, we expect $h+1$ inequivalent conformal field theories of this kind, for any given $h\geq 0$. Since there is no freedom to change these field theory data compatibly with the fivebrane configuration, according to the conjecture laid out at the end of section \ref{subsec:fracM2} we expect that $H^4_{tor}(L^{1,h+1,1,1},\bbZ)=\bbZ_{h+1}$. Note that in the case $h=1$ the 7-fold is $L^{1,2,1,1}\equiv Y^{1,2}(\bbC\bbP^2)$, which has $\bbZ_2$ torsion in the integral $H^4$ \cite{Martelli:2008rt,Benishti:2009ky}, in agreement with the expectation. We will check the conjecture for this family in the following.
 
The toric diagram of the second family of toric $CY_4$ cones with isolated singularity has vertices
\begin{equation}\label{toric_family_2}
\begin{split}
& a_h = (0,0,h),\quad a_{h+1}=(0,0,h+1),\quad b_0=(0,1,0),\\
& c_0=(1,1,0),\quad c_1=(1,1,1),\quad d_0=(0,1,0),
\end{split}
\end{equation}
with $h\neq -2,0$ (otherwise the singularity is not isolated). 
The GLSM has charges 
\be\label{GLSM_family_2}
\begin{tabular}{c c c c c c}
$a_h$ & $a_{h+1}$ &  $c_0$ & $c_1$ &$b_0$ & $d_0$ \\ \hline
 $-(h+1)$ & $h$  & $-1$ & $0$ &$1$ & $1$ \\
$1$ & $-1$ & $-1$ & $1$ &$0$ & $0$
\end{tabular}
\ee
$h\geq 1$ and $h\leq -3$ are equivalent, therefore we restrict our attention to the one parameter family $h\geq 1$ and to the exceptional case $h=-1$, which is the cone over $Q^{1,1,1}$.
For later convenience, we label the family of Sasaki-Einstein 7-folds which are the bases of these toric $CY_4$ cones as $Y^h$.

These manifolds span a one parameter subfamily of a six parameter family of smooth toric Sasaki-Einstein 7-folds whose metrics were found in \cite{Lu:2005sn}. The GLSM for the 6-parameter family of toric Calabi-Yau 4-folds with isolated singularities has $U(1)^2$ gauge group with charges $(m_1,m_2,-(m_1+m_2+2m_{56}),0,m_{56},m_{56})$, $(n_1,n_2,n_3,-(n_1+n_2+n_3),0,0)$ respectively.

The gauge theories for M2 branes probing $C(Y^h)$ have ABJM quivers, with CS levels $(h+1,-(h+1))$, a flavour pair coupled to $A_1$ and another flavour pair coupled to $A_2$, and superpotential \eqref{W_flavoured_KW} with $h_a=h_c=1$ and $h_b=h_d=0$. According to the discussion of sections \ref{subsec:brane_creation} and \ref{subsec:fracM2}, we expect $h+2$ inequivalent conformal field theories of this kind, for any given $h \geq 0$. Again there is no way of changing these field theory data compatibly with the fivebrane configuration in IIB, therefore according to the conjecture of section \ref{subsec:fracM2} we expect that $H^4_{tor}(Y^h,\bbZ)$ coincides with $\bbZ_{h+2}$ rather than simply containing it.  Notice that in the special case $h=-1$ that corresponds to $Q^{1,1,1}$, separated by the infinite class by the singular $SE_7$, the number of D3 branes created in a HW transition enhances to $n=2$.

The conjecture for the torsion part of the fourth cohomology based on brane physics and field theory arguments agrees with the algebraic topology computations performed in the literature for $Y^{1,2}(\mathbb{CP}^2)$ and $Q^{1,1,1}$, which both have $\bbZ_2$ torsion in the fourth integral cohomology \cite{Martelli:2008rt,Benishti:2009ky,Benishti:2010jn}. 
It would be nice to check explicitly the conjecture that the brane argument in type IIB saturates the order of the torsion part of the fourth cohomology group for the infinite classes of smooth Sasaki-Einstein 7-folds presented in this section. Unfortunately we are not aware of a general method to compute such cohomologies for toric Sasaki-Einstein 7-folds, and we have to resort to case by case computations. 
For the $L^{1,h+1,1,1}$ geometries, we can extend the computation of $H^4_{tor}$ provided for $L^{1,2,1,1}\equiv Y^{1,2}(\bbC\bbP^2)$ in \cite{Benishti:2009ky}, to which we refer the reader for more details and notation (that we will use in the next section), keeping track of some differences. 

Let us then consider the partial resolution of $C(L^{1,h+1,1,1})$ corresponding to positive FI parameter in the GLSM \eqref{GLSM_L_family}: it is a Calabi-Yau fourfold $\hat{X}_h$ which is the total space of $\calO(-1)\oplus \calO(-(h+1))\rightarrow \mathbb{WCP}^2_{[1,1,h]}$. $\hat{X}_h$ is contractible to its zero section (at $c_0=a_h=0$), which is a copy of $\mathbb{WCP}^2_{[1,1,h]}$, therefore it has only nonvanishing integral cohomologies $H^0(\hat{X}_h,\bbZ)=H^2(\hat{X}_h,\bbZ)=H^4(\hat{X}_h,\bbZ)=\bbZ$ \cite{Kawasaki}. Note that $\hat{X}_h$ has a $\bbZ_h$ orbifold singularity (if $h>1$) at the point $p$ determined by $c_0=a_h=b_0=d_0=0$, that we can excise to get a smooth $X_h$. The cohomologies written above for $\hat{X}_h$ extend to the smooth $X_h$ via the excision theorem: $H^0(X_h,\bbZ)=H^2(X_h,\bbZ)=H^4(X_h,\bbZ)=\bbZ$.
The removal of the point $p$ can be thought of physically as due to the backreaction of M2 branes placed at $p$, which sends the point to infinity creating an $AdS_4\times Y_{IR}$ throat \cite{Benishti:2009ky}. Sending the removed point to infinity does not change the topology of $X_h$. The boundary of $X_h$ can then be thought of as $\partial X_h=Y_{UV}\cup Y_{IR}$ (the union is disjoint), where $Y_{UV}=L^{1,h+1,1,1}$ and $Y_{UV}=S^7/\bbZ_h$, whereas $\partial \hat{X}_h=Y_{UV}$ (but there is a leftover singularity). 
By the Thom isomorphism, $H^4(X_h, \partial X_h,\bbZ) = H^6(X_h, \partial X_h,\bbZ)= H^8(X_h, \partial X_h,\bbZ)=\bbZ$, where the generator of $H^4(X_h, \partial X_h,\bbZ)$ is the Thom class (and similarly for $\hat{X}_h$).
The image of the generator in the map from $H^4(X_h, \partial X_h,\bbZ)=H^4(\hat{X}_h, L^{1,h+1,1,1},\bbZ)$ to $H^4(X_h,\bbZ)=H^4(\hat{X}_h,\bbZ)=H^4(\mathbb{WCP}^2_{[1,1,h]},\bbZ)=\bbZ$ is then the Euler class of the $\calO(-1)\oplus \calO(-(h+1))$ bundle, which is the element $h(h+1)\in \bbZ=H^4(\mathbb{WCP}^2_{[1,1,h]},\bbZ)$. Here the additional $h$ factor compared  to \cite{Benishti:2009ky} comes from the self-intersection number $h$ of the generator of $H_2(\mathbb{WCP}^2_{[1,1,h]},\bbZ)$ \cite{Kawasaki}.
By means of a long exact sequence similar to (5.4) of \cite{Benishti:2009ky}, it  follows that $H^4(\partial X_h,\bbZ)=\bbZ_{h(h+1)}=\bbZ_h\oplus \bbZ_{h+1}$. Since $H^4(\partial X_h,\bbZ)=H^4(L^{1,h+1,1,1},\bbZ)\oplus H^4(S^7/\bbZ_h,\bbZ)$ and $H^4(S^7/\bbZ_h,\bbZ)=\bbZ_h$, we conclude that $H^4(L^{1,h+1,1,1},\bbZ)=\bbZ_{h+1}$, as expected.

%
%
%

It would be interesting to perform a similar check for the geometries $Y_h$, for which we have not been able to evaluate the fourth cohomologies. 
In the next section we will study partial resolutions and $G$-fluxes in more detail from the type IIB perspective.

For future reference, in appendix \ref{app:volumes} we collect details about the volumes of the smooth Sasaki-Einstein manifolds and their supersymmetric 5-cycles, from which the superconformal R-charges (equal to the conformal dimensions) of matter fields can be computed.

\section{Fractional M2 branes and partial resolutions}\label{sec:Higgsing}

The beautiful interplay of torsion G-fluxes in the M-theory backgrounds and partial resolutions of Calabi-Yau fourfold singularities was investigated in detail by Benishti, He and Sparks in \cite{Benishti:2009ky}. We now review some of the salient aspects of their general analysis, adapting the discussion to field theories that may include fundamental flavours.
 
Given a toric $CY_4$ cone $\bar{X}$ and a partial Calabi-Yau resolution thereof $\pi:\,\hat{X}\to\bar{X}$, we can consider placing a stack of $N$ regular M2 branes at a (possibly residually singular) point $p\in \hat{X}$.
Let $C(Y_{IR})$ be the tangent space at $p$.
After including the backreaction of the $N$ M2 branes, $p$ is sent to infinity inside a throat. The resulting background is a warped product $\bbR^{1,2}\,\times_w\, X$, where $X=\hat{X}\setminus p$, with two disconnected boundaries where the solution asymptotes to $AdS_4\times Y_{UV}$ and $AdS_4\times Y_{IR}$ respectively.
In the dual field theory, the partial resolution is interpreted as due to Fayet-Iliopoulos terms or real mass terms for flavour fields.
This relevant deformation induces an RG flow driving the UV field theory, holographically dual to $AdS_4\times Y_{UV}$, to a different infrared fixed point, holographically dual to $AdS_4\times Y_{IR}$.%

In the presence of torsion G-flux in $AdS_4\times Y_{UV}$, several scenarios might occur \cite{Benishti:2009ky}. First, it is possible that the torsion G-flux cannot be extended to the supergravity solution for the holographic RG-flow corresponding to a partial resolution. In that event, the putative supersymmetric RG flow of the dual field theory does not exist.
Second, if instead the torsion G-flux can be extended, the extension may be ambiguous and asymptote to various torsion classes in $H^4(Y_{IR},\bbZ)$. If so, there are several RG flows driving a given UV fixed point to distinct IR fixed points labelled by different ranks of the gauge groups (but the same quiver and superpotential).

In this section we would like to analyse the interplay between partial resolution and torsion $G$-fluxes by means of the dual type IIB brane constructions of the models under investigation. Rather than delving into a general discussion, we begin focusing on the example $Y_{UV}=Y^{1,2}(\mathbb{CP}^2)$, which was studied in detail in \cite{Benishti:2009ky}. The generalisation is straightforward and is left as an exercise for the interested readers.

\begin{figure}
 \begin{minipage}[t]{7cm}
   \centering
\includegraphics[width=6cm]{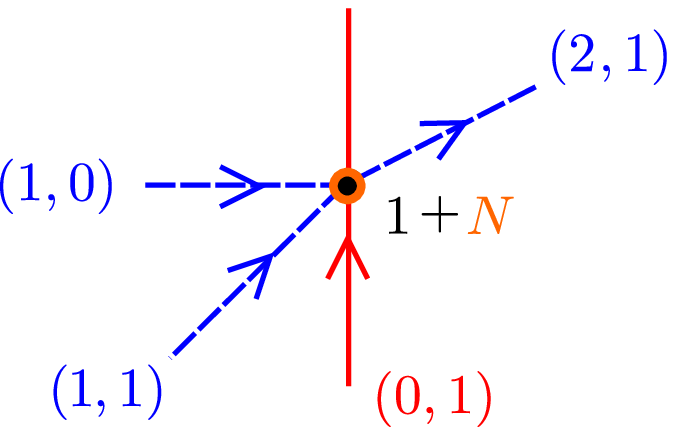}
\caption{\small 5-brane configuration for $C(Y^{1,2}(\mathbb{CP}^2))$ with $N$ regular M2 branes (in orange) and 1 fractional M2 brane (in black).}\label{fig:Y21_5branes_sing}
 \end{minipage}
 \ \hspace{2mm} \hspace{3mm} \
 \begin{minipage}[t]{7cm}
  \centering
\includegraphics[width=6cm]{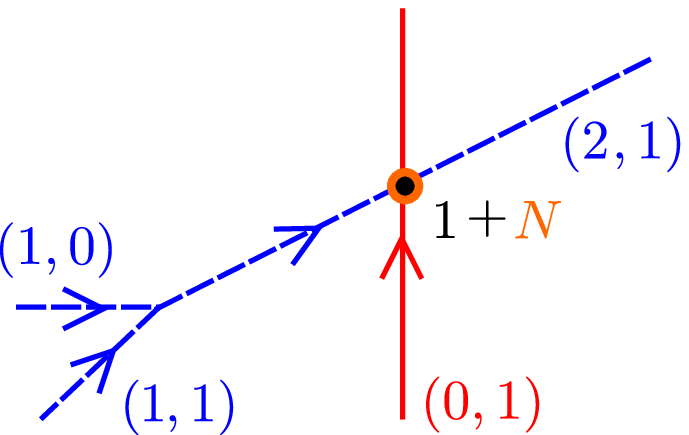}
\caption{\small 5-brane configuration for the partial resolution of $C(Y^{1,2}(\mathbb{CP}^2))$ to $\bbC^4/\bbZ_2$ with $N$ regular M2 branes (in orange) and 1 fractional M2 brane (in black).}\label{fig:Y21_5branes_resol_C4Z2}
 \end{minipage}
\end{figure}

Let us then consider the type IIB brane construction (in the $37$ plane) for $N$ regular plus $1$ fractional M2 branes at $C(Y^{1,2}(\mathbb{CP}^2))$, see fig. \ref{fig:Y21_5branes_sing}: regular M2 branes are dual to D3 branes wrapping the $x^6$ circle, which are depicted in orange; fractional M2 branes are dual to D3 branes suspended between the two fivebranes along an interval of the $x^6$ circle, which are depicted in black.  
The 3d field theory is a $U(N+1)_{3/2}\times U(N)_{-3/2}$ theory with a single flavour pair coupled to $A_1$ as in \eqref{W_flavoured_KW} with $h_a=1$ and $h_b=h_c=h_d=0$.

We are interested in partial resolutions of the singularity, and we focus on the situation where all the $N$ regular M2 branes are left at the residual singularity. As for the fate of the fractional M2 branes, they can either remain at the residual singularity during the partial resolution, or be removed from it in the process.

We first study the resolution to $\bbC^4/\bbZ_2$, which has a dual interpretation as a real mass term for the flavours. The dual global deformation of the type IIB fivebrane configuration is shown in fig. \ref{fig:Y21_5branes_resol_C4Z2}. As the fivebrane junction is removed, the D3 brane on the interval is forced to remain suspended between the $(2,1)$5 and the NS5 brane if we require supersymmetry. The low energy field theory is therefore the $U(N+1)_{2}\times U(N)_{-2}$ ABJ theory. The dual supergravity solution has one unit of torsion G-flux  in $Y_{IR}=S^7/\bbZ_2$.

Similarly, the absence of torsion G-flux in $Y_{UV}$ implies the absence of torsion G-flux in $Y_{IR}$. This is easy to understand: if one starts with only D3 wrapping the circle, one ends up with only D3 wrapping the circle. Alternatively, by a Hanany-Witten transition, if one starts with the maximal number of D3 branes on an interval allowed by the $s$-rule, one also ends up with the maximal number.

\begin{figure}
 \begin{minipage}[t]{7cm}
   \centering
\includegraphics[width=6cm]{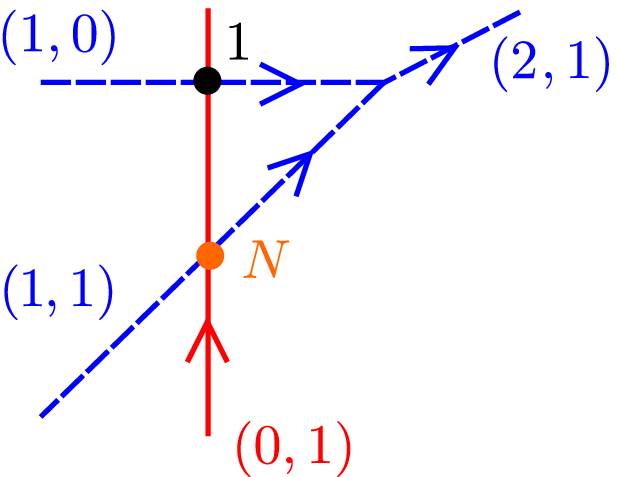}
\caption{\small 5-brane configuration for a resolution of $C(Y^{1,2}(\mathbb{CP}^2))$ with $N$ regular M2 branes  and 1 fractional M2 brane to $\bbC^4$.}\label{fig:Y21_5branes_resol_C4N} 
 \end{minipage}
 \ \hspace{2mm} \hspace{3mm} \
 \begin{minipage}[t]{7cm}
  \centering
\includegraphics[width=6cm]{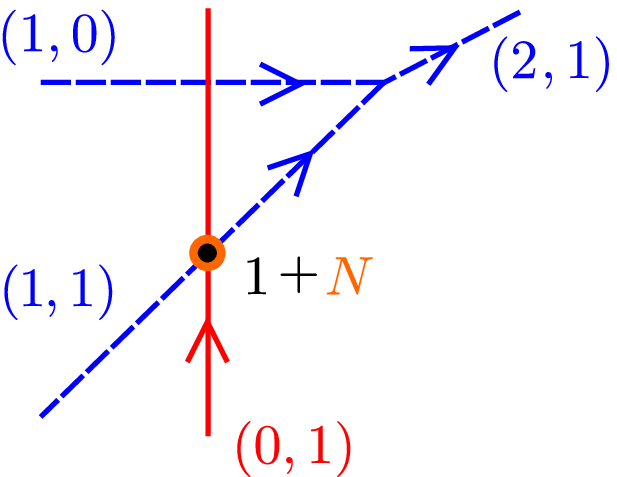}
\caption{\small 5-brane configuration for a resolution of $C(Y^{1,2}(\mathbb{CP}^2))$ with $N$ regular M2 branes  and 1 fractional M2 brane to $\bbC^4$.}\label{fig:Y21_5branes_resol_C4Np1} 
\end{minipage}
\end{figure}

The other resolution of $C(Y^{1,2}(\mathbb{CP}^2))$ is to $\bbC^4$. In type IIB brane terms, there are several options for the leftover fivebranes, leading to apparently different 3d field theories which however are mirror symmetric. We focus for definiteness on the situation where we are left with a $(1,1)5$ and an NS5, namely the ABJ(M) setup. The dual interpretation of this resolution is again a real mass term for the flavours, but of opposite sign compared to the previous case.
The D3 on the interval can either move to infinity or not as the resolution parameter is sent to infinity, as shown in figures \ref{fig:Y21_5branes_resol_C4N} and \ref{fig:Y21_5branes_resol_C4Np1} respectively. The resulting low energy conformal field theories are the $U(N)_1\times U(N)_{-1}$ ABJM model or the $U(N+1)_1\times U(N)_{-1}$ ABJ model, which are equivalent \cite{Aharony:2008ug}.
The supergravity solution has (necessarily) no torsion G-flux in $Y_{IR}=S^7$.
These results, which are straightforward to derive using branes in type IIB string theory, agree with the more complicated cohomological analysis of \cite{Benishti:2009ky}.

\begin{figure}
 \begin{minipage}[t]{7cm}
   \centering
\includegraphics[width=6cm]{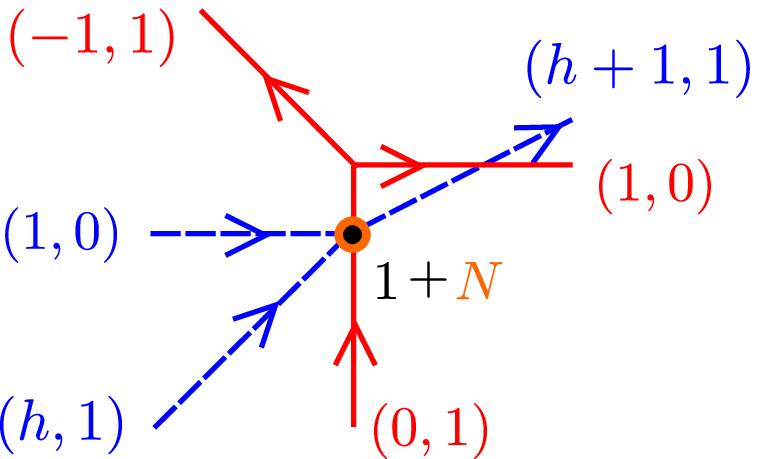}
\caption{\small 5-brane configuration for a resolution of $C(Y^h)$ to $C(L^{1,h+1,1,1})$ with $N$ regular M2 branes and 1 fractional M2 brane both at the unresolved and the partially resolved singularity.}\label{fig:Yh_sing_Np1_resol} 
 \end{minipage}
 \ \hspace{2mm} \hspace{3mm} \
 \begin{minipage}[t]{7cm}
  \centering
\includegraphics[width=6cm]{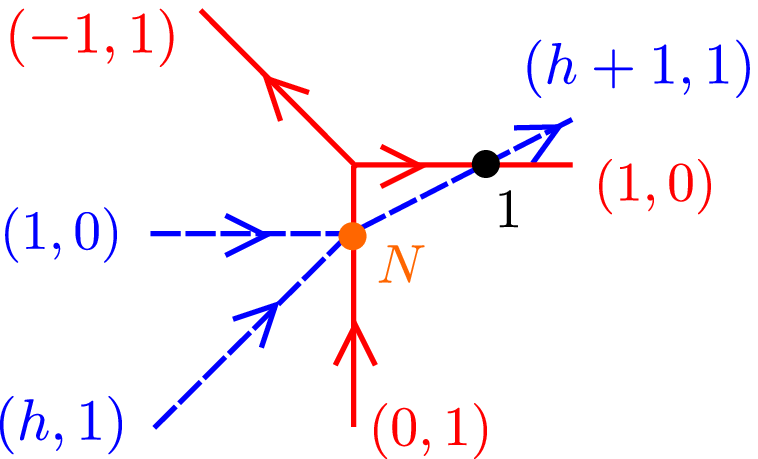}
\caption{\small 5-brane configuration for a resolution of $C(Y^h)$ to $C(L^{1,h+1,1,1})$ with $N$ regular M2 branes and 1 fractional M2 brane at the unresolved singularity, and only $N$ regular M2 branes at the partially resolved singularity.}\label{fig:Yh_sing_N_resol} 
\end{minipage}
\end{figure}

As an example of a holographic RG flow where torsion $G$-flux in $Y_{UV}$ can be extended in several ways, leading to IR backgrounds with different torsion $G$-fluxes in the same $Y_{IR}$, let us take $Y_{UV}=Y^h$ and $Y_{IR}=L^{1,h+1,1,1}$, and choose for instance one unit of torsion flux in $Y_{UV}$. Remember that according to section \ref{sec:smooth}, $H^4_{tor}(Y^h,\bbZ)=\bbZ_{h+2}$ and $H^4_{tor}(L^{1,h+1,1,1},\bbZ)=\bbZ_{h+1}$.
Translating the partial resolution and the fractional M2 branes into the type IIB brane construction, we see in figures \ref{fig:Yh_sing_Np1_resol}-\ref{fig:Yh_sing_N_resol} that the partial resolution can result either in one unit of torsion $G$-flux in $Y_{IR}$ or in no torsion.

Even though we have concentrated on a couple of examples, we hope it is clear that the philosophy can easily be applied to the entire class of flavoured ABJ(M) models studied in this paper. 

Concerning potential obstructions to extending the torsion $G$-flux in $Y_{UV}$ to a $G$-flux on $X$, we see no trace of it in our analysis, which applies to the geometries arising as moduli spaces of flavoured ABJ(M) models. The total number $n$ of D3 branes created in a HW transition in the type IIB dual does not change under finite deformations of the webs, although some D3 branes may be forced to follow the deformation. It can stay constant or decrease if a deformation parameter is sent to infinity. Consequently, given a configuration of D3 branes that satisfies the $s$-rule before the deformation, it is always possible to accommodate the $s$-rule after the deformation as well. This result seems to imply that there is always a supersymmetric RG flow between a UV and an IR supersymmetric fixed point in our field theories when real masses or FI terms are introduced.

Note that partial resolutions may be used to set a lower bound on $n$ before computing it, if $n$ is known for the partially resolved singularity. For instance, we can consider partial resolutions to ABJM singularities $\bbC^4/\bbZ_{|k_{max}|}$, where $|k_{max}|\geq 0$ is maximised. Thinking about the dual fiald theories, we see that $|k_{max}|=|k|+\frac{1}{2}(N_L+N_R)$, where $k$, $N_L$ and $N_R$ as in \eqref{N_L}-\eqref{CS_level_from_h} refer to the original flavoured ABJM field theory. In classes 1 and 4 of section \ref{subsec:brane_creation}, $|k_{max}|=n$: there is a partial resolution to $\bbC^4/\bbZ_n$, and the result for $n$ should not be surprising. On the other hand, in classes 2 and 3, which require flavours in non-real representations of the gauge groups, $|k_{max}|\leq n=\max\{N_L,N_R\}$, and the inequality is saturated only at the boundary of these classes, where $h$ or $\hat{h}$ vanish. Therefore the value of $n$ cannot be guessed just by considering partial resolutions. 
An example of this is $C(Q^{1,1,1})$, in which case any partial resolution reduces $n$ from 2 to 1.

\begin{figure}[t]
\begin{center}
\includegraphics[width=6cm]{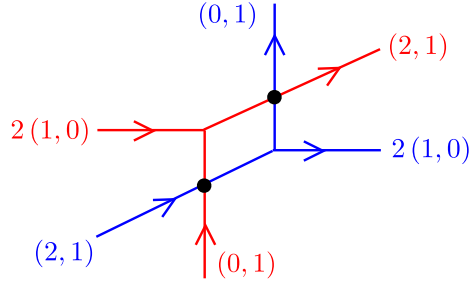}
\caption{\small 5-brane configuration for a resolution of $C(Q^{1,1,1}/\bbZ_2)$.}\label{fig:resol_Q111Z2}
\end{center}
\end{figure}

Finally, one can also enjoy the possibility of displacing singularities by a partial resolution, and partitioning the (regular and fractional) M2 branes among the daughter singularities in the process.
In the supergravity solution, one would have multiple `infrared' $AdS_4\times Y_{IR,i}$ throats, possibly with torsion $G$-fluxes even when there is no torsion $G$-flux in $Y_{UV}$. In the dual field theory, there would be several SCFT's coupled by irrelevant operators.%
\footnote{Similar aspects have been studied in the context of the $AdS_5/CFT_4$ correspondence  \cite{GarciaEtxebarria:2006aq}.}
A novelty of $AdS_4/CFT_3$ is that these SCFTs may have different ranks.
For instance, we may partially resolve $C(Q^{1,1,1}/\bbZ_2)$, where the $Z_2$ action has fixed points as in \cite{Benini:2009qs}, as dually depicted in fig. \ref{fig:resol_Q111Z2}, and choose to place all the colour branes at the two leftover $\bbC^4/\bbZ_2$ isolated singularities as labelled by the black dots.%
Suppose that the starting point has only $N$ regular M2 branes at $C(Q^{1,1,1}/\bbZ_2)$. Then we may end up with $N_1$ regular M2 at a $\bbC^4/\bbZ_2$ singularity and $N-N_1$ regular M2 at the other $\bbC^4/\bbZ_2$ singularity. But we could also fractionate a regular M2 into two fractional M2 branes of different kinds, and separate the two fractional branes: then we would end up with $N_1$ regular M2 together with a fractional M2 brane of one type at a $\bbC^4/\bbZ_2$ singularity, and 
$N-N_1-1$ regular M2 together with a fractional M2 brane of the other type at the other $\bbC^4/\bbZ_2$ singularity. The dual type IIB brane configuration still satisfies the $s$-rule after the deformation of the brane webs. In this case, the two conformal field theories which are coupled only by massive fields have gauge groups and CS levels $U(N_1+1)_1\times U(N_1)_{-1}$ and $U(N-N_1-1)_1\times U(N-N_1)_{-1}$.


\section{Conclusions}\label{sec:conclusions}

In this article, we presented type IIB brane realisations of Yang-Mills-Chern-Simons ABJ(M) theories with fundamental and antifundamental flavours, whose infrared fixed points describe 3d theories on M2 branes at certain toric Calabi-Yau fourfolds relevant to $AdS_4/CFT_3$.

The existence of a type IIB dual setup of D3 branes suspended between fivebranes is limited to M2 branes at a subset of all toric Calabi-Yau fourfolds. On the other hand, this duality is very powerful when available, having the potential of translating difficult problems, such as understanding duality cascades  and identifying the conformal window of the field theory, into simpler ones that can be answered by inspection of fivebrane web diagrams. 

In this paper we focused our attention on fractional M2 branes which arise as the uplift of D4 branes wrapping the vanishing 2-cycle of the conifold in IIA and do not spoil the $AdS_4$ near horizon geometry. Their manifestation in the 11d gravity dual $AdS_4\times Y_7$ background is a torsion $G$-flux in $Y_7$.
Such fractional M2 branes are related to the different 3d superconformal field theories with given quiver diagram, superpotential, CS levels and minimal common rank $N$ of the gauge groups.
Duality maps these fractional M2 branes into D3 branes stretched along an interval between two fivebrane webs on a circle. The maximal number of such branes that can be added to $N$ D3 branes wrapping the circle (dual to regular M2 branes) compatibly with a superconformal fixed point of the 3d gauge theory is determined by a basic rule in brane physics, the $s$-rule of \cite{Hanany:1996ie,Kitao:1998mf}, which we implemented for fivebrane systems involving junctions. 

We have identified the superconformal field theories dual to $AdS_4\times Y_7$ backgrounds of M-theory without and with torsion $G$-fluxes for the smooth geometries $Y_7=Q^{1,1,1},\; L^{1,h+1,1,1}$ ($h\geq 1$), matching the number of inequivalent superconformal field theories with given quiver diagram, superpotential, and common rank $N$ of the gauge groups that we found in the type IIB dual, with the order of the fourth cohomology group of the Sasaki-Einstein manifolds. 

From our point of view, the type IIB dual construction has additional advantages. First, it confirms the proposal of \cite{Benini:2009qs,Jafferis:2009th} that the so called three-dimensional quiver gauge theories without four-dimensional parent that received wide attention in the literature should be replaced by ordinary quiver gauge theories coupled to fundamental and antifundamental flavours, manifesting the connection between the flavoured ABJ(M) field theory on M2 branes and the toric $CY_4$ geometry that they probe.
 Second, it provides a simplified arena for interpreting partial resolutions of the Calabi-Yau fourfold in the dual three dimensional gauge thory, by mapping them to web deformations of the fivebrane system.
As we explained, in the context of the flavoured ABJ(M) models the proposed three-dimensional field theories for M2 branes allow a description of all the partial resolutions of the Calabi-Yau singularities, either as real mass terms for fundamental flavours or as Fayet-Iliopoulos terms which induce Higgsing.
We also showed how the interplay of partial resolutions and $G$-fluxes first studied in some particular cases in \cite{Benishti:2009ky} can be easilyy analysed by means of the type IIB brane cartoon.

On the other hand, it would be of great interest to understand how the addition of new points in the toric diagram can be accounted for by introducing new degrees of freedom in the 3d gauge theory. This program goes under the name of \emph{unHiggsing}%
\footnote{This name, inherited from the $AdS_5/CFT_4$ correspondence, is misleading in the $AdS_4/CFT_3$ correspondence since partial resolutions are not necessarily described as Higgsings in this context.}
 and is one of the major open problems in the $AdS_4/CFT_3$ correspondence for M2 branes at toric $CY_4$ singularities.

The type IIB dual construction looks promising in this respect. The relation between toric diagrams and fivebrane webs discussed in section \ref{sec:webs-toric} for our models suggests a natural generalisation to two arbitrary 5-brane systems (still at an angle in the 45-89 plane) in type IIB: this is achieved by pulling in, from infinity in the 37 plane, fivebrane subwebs which are suitably oriented in the 45-89 plane.%
\footnote{See also section 4 of \cite{Aharony:1997ju} for a related discussion of non-elliptic models of D3 branes suspended on an interval between two generic fivebrane webs.}
This operation adds points to the layers of the toric diagram of the fourfold, so that we end up with a toric diagram which is still made of two vertical layers, each one being the toric diagram of the toric $CY_3$ dual to a single $(p,q)5$ web. Note that the restriction to two layers rules out fourfolds with exceptional divisors.

In the literature there are proposals for 3d field theories on M2 branes at only some of these toric $CY_4$ cones. Even when available, they are generically plagued by difficulties in accounting for some partial resolutions. However, such partial resolutions correspond to web deformations that are manifest in the type IIB brane picture. They occur when two fivebrane subwebs at equilibrium (with the same orientation in the 45-89 plane) are moved to the same position in the $x^6$ circle, so that they intersect and can undergo a web deformation. From the point of view of the dual ultraviolet 3d gauge theory, this is a limit of infinite Yang-Mills coupling, where new degrees of freedom appear.
We hope to report progress in understanding the 3d field theories dual to generic two-fivebranes systems and the field-theoretical interpretation of partial resolutions in the near future.


\vspace{15pt}

\section*{Acknowledgments}

It is a pleasure to thank Ofer Aharony, Francesco Benini, Oren Bergman and especially Cyril Closset for valuable discussions at various stages of this project, and Jarah Evslin, Dario Martelli and Roberto Valandro for discussions on topological aspects.
The work of the author is supported in part by the Israeli Science Foundation centre of excellence, by the Deutsch-Israelische Projektkooperation (DIP), by the US-Israel Binational Science Foundation (BSF), and by the German-Israeli Foundation (GIF).

\vspace{15pt}

\appendix
\section{Volumes of smooth $SE_7$ and superconformal R-charges}\label{app:volumes}

In this appendix we list the outcome of some computations of volumes for the smooth Sasaki-Einstein 7-folds introduced in section \ref{sec:smooth} and their calibrated 5-cycles (the cones over which are toric divisors), obtained by volume minimisation \cite{Martelli:2005tp} following the appendix of \cite{Hanany:2008fj}. Knowledge of the volumes of calibrated 5-cycles, on which M5 branes can be wrapped supersymmetrically, allows us to compute the superconformal R-charges of the matter fields, which are equal to their conformal dimensions. 

\subsection{$L^{1,h+1,1,1}$, $h\geq 1$}

Thanks to symmetries, the Reeb vector field of the Sasakian 7-fold can be written as $b=(4,x,x,y)$, where 
\begin{equation}\label{range_L}
0<x<4\,,\quad y>0\,,\quad 2h(2-x)<y<(h+1)(4-x)\;.
\end{equation} 
The volume is 
\begin{equation}\label{volumeL}
\Vol(L^{1,h+1,1,1})=\frac{\pi^4}{3} \cdot \frac{h(h+1)^2 x^2 + y[4(h+1)-y]}{x^2 y \left[y-(1+h)(4-x)\right]^2[y-2h(2-x)]}\;.
\end{equation}
Minimisation of this volume in the range \eqref{volumeL} determines the volume of the Sasaki-Einstein 7-fold. From the computation of volumes of supersymmetric 5-cycles, we can extract the R-charges of the perfect matchings 
\begin{align}
R[a_h]&= \frac{1}{2} \, \frac{y [(h+1)(4-x)-y]^2}{h(h+1)^2 x^2 + y[4(h+1)-y]}\\
R[a_{h+1}]&= \frac{1}{2} \, \frac{y [4(h+1)-y][y-2h(2-x)]}{h(h+1)^2 x^2 + y[4(h+1)-y]}\\
R[c_0]&= \frac{1}{2}\,\frac{(h+1)^2 x^2 [y-2h(2-x)]}{ h(h+1)^2 x^2 + y[4(h+1)-y]}\\
R[b_0]= R[d_0] &= \frac{1}{2} \, \frac{x [(h+1)(4-x)-y][y+h(h+1)x]}{h(h+1)^2 x^2 + y[4(h+1)-y]}
\end{align}
where again it is understood that the values of $x$ and $y$ that minimise \eqref{volumeL} have to be inserted.
We recall that bifundamentals and diagonal monopole operators are related to the perfect matchings as follows \cite{Benini:2009qs}: 
\begin{equation}\label{matchings-fields_L}
T=a_h\,,\quad \tilde{T}=a_{h+1}\,,\quad A_1=a_h a_{h+1}\,,\quad B_1=b_0\,,\quad A_2=c_0\,,\quad B_2=d_0\,.
\end{equation}
Under the assumption that the entire superpotential \eqref{W_flavoured_KW} is marginal, we can  determine the superconformal R-charge of the flavour fields: $R[p_1]=R[q_1]=1-\frac{1}{2}R[A_1]$.

Unfortunately, we have not been able to find a compact analytic form for the values of $x$ and $y$ that determine the Sasaki-Einstein manifold as a function of $h$, but the minimisation can be readily performed numerically.

\subsubsection{$h=1$: $L^{1,2,1,1}=Y^{1,2}(\mathbb{CP}^2)$}

For $h=1$, which is $L^{1,2,1,1}=Y^{1,2}(\mathbb{CP}^2)$, the values of $(x,y)$ which minimise the volume are
\begin{equation}
x=\frac{1}{3}\left(1+23\, c^{-1/3} -c^{1/3}\right)\;,\qquad y=4-x\;,\qquad c\equiv -181+24\sqrt{78}\;.
\end{equation}
The volume of the Sasaki-Einstein manifold is 
\begin{equation}\label{volumeY12}
\Vol(Y^{1,2}(\mathbb{CP}^2))=\frac{\pi^4}{3}\,\frac{16+3x^2}{x^3(4-x)^3}\simeq 13.3916
\end{equation}
which agrees with \cite{Martelli:2008rt}, up to a prefactor of 4 which is perhaps a typo. The volumes of supersymmetric 5-cycles are:
\begin{equation}
\begin{split}
\Vol(\Sigma_{a_1})&=\pi^3\,\frac{1}{x^3}\simeq 6.0075\\
\Vol(\Sigma_{a_2})=\Vol(\Sigma_{b_0})=\Vol(\Sigma_{d_0})&=\pi^3\,\frac{4+x}{x^2(4-x)^2}\simeq 11.5224\\
\Vol(\Sigma_{c_0})&=\pi^3\,\frac{4}{(4-x)^3}\simeq 10.5774
\end{split}
\end{equation}
The superconformal R-charges of the matter fields are:
\begin{equation}
\begin{split}
R[T]&=\frac{1}{2}\,\frac{(4-x)^3}{16+3x^2}\simeq 0.2349\\
R[\tilde{T}]=R[B_i]&=\frac{1}{2}\,\frac{x(16-x^2)}{16+3x^2}\simeq 0.4505\\
R[A_1]=R[T]+R[\tilde{T}]&= \frac{(4-x)(8-2x+x^2)}{16+3x^2}\simeq 0.6854\\
R[A_2]&=\frac{2x^3}{16+3x^2}\simeq 0.4136\\
R[p_1]=R[q_1]=1-\frac{1}{2}R[A_1]&=\frac{x(16+x^2)}{16+3x^2}\simeq 0.6573
\end{split}
\end{equation}

\subsubsection{$h\to\infty$}
In the $h\to\infty$ limit, the values of $(x,y)$ which minimise the volume of $L^{1,h+1,1,1}$ are 
\begin{equation}
\begin{split}
x & = 2-\frac{3}{8}\,h^{-1}+\frac{7}{128}\,h^{-2}+\calO(h^{-3})\\
y & = h\left( 1+\frac{5}{4}\,h^{-1}+\frac{5}{32}\,h^{-2}+\calO(h^{-3})\right)\;.
\end{split}
\end{equation}
The volume of the Sasaki-Einstein manifold is 
\begin{equation}
\Vol(L^{1,h+1,1,1})=\frac{\pi^4}{3h} \,\left( 1-\frac{5}{4}\,h^{-1}+\frac{85}{64}\,h^{-2}+\calO(h^{-3})\right)\;,
\end{equation}
the volumes of calibrated 5-cycles are
\begin{equation}
\begin{split}
\Vol(\Sigma_{a_h})&=\frac{\pi^3}{4h}\,\left( 1-\frac{1}{8}\,h^{-1}-\frac{39}{256}\,h^{-2}+\calO(h^{-3})\right)  \\
\Vol(\Sigma_{a_{h+1}})&=\frac{3\pi^3}{4h}\,\left( 1-\frac{23}{24}\,h^{-1}+\frac{695}{768}\,h^{-2}+\calO(h^{-3})\right)  \\
\Vol(\Sigma_{b_0})=\Vol(\Sigma_{d_0})&=\frac{\pi^3}{h}\,\left( 1-\frac{11}{8}\,h^{-1}+\frac{191}{128}\,h^{-2}+\calO(h^{-3})\right)  \\
\Vol(\Sigma_{c_0})&=\frac{\pi^3}{h}\,\left( 1-\frac{3}{2}\,h^{-1}+\frac{27}{16}\,h^{-2}+\calO(h^{-3})\right)  \;, 
\end{split}
\end{equation}
and the superconformal R-charges are
\begin{equation}
\begin{split}
R[T]&=  \frac{1}{8}\,\left(1+\frac{9}{8}\,h^{-1}-\frac{19}{256}\,h^{-2}+\calO(h^{-3})\right)\\
R[\tilde{T}]&= \frac{3}{8}\,\left(1+\frac{7}{24}\,h^{-1}-\frac{45}{768}\,h^{-2}+\calO(h^{-3})\right)\\
R[B_i] &= \frac{1}{2}\,\left(1-\frac{1}{8}\,h^{-1}+\frac{1}{128}\,h^{-2}+\calO(h^{-3})\right)\\
R[A_1]&= \frac{1}{2}\,\left(1+\frac{1}{2}\,h^{-1}-\frac{1}{16}\,h^{-2}+\calO(h^{-3})\right)\\
R[A_2]&= \frac{1}{2}\,\left(1-\frac{1}{4}\,h^{-1}+\frac{3}{64}\,h^{-2}+\calO(h^{-3})\right)\\
R[p_1]=R[q_1]&= \frac{3}{4}\,\left(1-\frac{1}{6}\,h^{-1}+\frac{1}{48}\,h^{-2}+\calO(h^{-3})\right)\;.
\end{split}
\end{equation}


\subsection{$Y^h$, $h\geq 1$}

In this case, due to symmetries, there are only three independent volumes of 5-cycles, and the Reeb vector can be taken as $b=(4,2,2,y)$, with $0<y<2(h+2)$. The volume of the Sasakian manifold with this Reeb vector is
\begin{equation}\label{volumeY}
\Vol(Y^h)=\frac{\pi^4}{6} \, \frac{2h(h+2)^2 +y[4(h+2)-y] }{y^2 [2( h + 2) - y]^2}\;.
\end{equation}
The volumes of calibrated 5-cycles are 
\begin{equation}
\begin{split}
\Vol(\Sigma_{a_h})=Vol(\Sigma_{c_0})&= \frac{\pi^3}{4y} \\
\Vol(\Sigma_{a_{h+1}})=Vol(\Sigma_{c_1})&= \frac{\pi^3}{4}\,\frac{4(h+2)-y}{[2( h + 2) - y]^2} \\
\Vol(\Sigma_{b_0})=Vol(\Sigma_{d_0})&= \pi^3\,\frac{h(h+2)+y}{y^2 [2( h + 2) - y] } \;.
\end{split}
\end{equation}
The superconformal R-charges are 
\begin{equation}
\begin{split}
R[A_i]&= \frac{(h+2)^2 y}{2h(h+2)^2 +y[4(h+2)-y]}\\
R[B_j]&= \frac{1}{4}\,\frac{[2( h + 2) - y](h^2+2h+y)}{2h(h+2)^2+y[4(h+2)-y]}\\
R[T]&= \frac{1}{2}\,\frac{[2( h + 2) - y]^2 y}{2h(h+2)^2 +y[4(h+2)-y]}\\
R[\tilde{T}]&= \frac{1}{2}\,\frac{[4( h + 2) - y]y^2}{2h(h+2)^2 +y[4(h+2)-y]}\\
R[p_i]=R[q_i]&=\frac{1}{2}\,\frac{4 h (h+2)^2 + 8 (h+2) y - (h+2)^2 y - 2 y^2}{2h(h+2)^2+y[4(h+2)-y]}\;.
\end{split}
\end{equation}
Minimising the volume \eqref{volumeY} amounts to solving a cubic equation. The solution is 
\begin{equation}\label{y_Y}
y = 2(h+2) - \frac{4}{\sqrt{3}} \,(h+2)^{3/2} \,\sin\left[\frac{1}{3} \arcsin\left(\frac{3\sqrt{3}}{4\sqrt{h+2}} \right)\right]\;,
\end{equation}
which can be plugged in the previous formulae to give the volumes of the Sasaki-Einstein $Y^h$, the volumes of their supersymetric 5-cycles, and the superconformal R-charges of quiver fields.


\end{document}